\documentclass{emulateapj}

\newcommand{\myemail}{t.hutchinson@utah.edu}
\newcommand{\redmonster}{\texttt{redmonster}}

\slugcomment{Accepted by AJ -- \today}

\shorttitle{redmonster}
\shortauthors{Hutchinson et al.}

\usepackage{float}

\begin{document}

\title{Redshift Measurement and Spectral Classification\\
	for \MakeLowercase{e}BOSS Galaxies with the \texttt{redmonster} Software}

\author{Timothy A. Hutchinson\altaffilmark{1},
	Adam S. Bolton\altaffilmark{1,2},
	Kyle S. Dawson\altaffilmark{1},
	Carlos Allende Prieto\altaffilmark{3,4},
	Stephen Bailey\altaffilmark{5},
	Julian E. Bautista\altaffilmark{1},
	Joel R. Brownstein\altaffilmark{1},
	Charlie Conroy\altaffilmark{6},
	Julien Guy\altaffilmark{7},
	Adam D. Myers\altaffilmark{8},
	Jeffrey A. Newman\altaffilmark{9},
	Abhishek Prakash\altaffilmark{9},
	Aurelio Carnero-Rosell\altaffilmark{10,11},
	Hee-Jong Seo\altaffilmark{12},
	Rita Tojeiro\altaffilmark{13},
	M. Vivek\altaffilmark{1},
	Guangtun Ben Zhu\altaffilmark{14}
}

\altaffiltext{1}{Department of Physics and Astronomy, University of Utah,
    Salt Lake City, UT 84112, USA (\texttt{\myemail})}

\altaffiltext{2}{National Optical Astronomy Observatory, 950 N. Cherry Ave.,
	Tucson, AZ 85719, USA}

\altaffiltext{3}{Instituto de Astrof\'{i}sica de Canarias, V\'{i}a L\'{a}ctea, 38205
	La Laguna, Tenerife, Spain}

\altaffiltext{4}{Universidad de La Laguna, Departamento de Astrof\'{i}sica,
	38206 La Laguna, Tenerife, Spain}

\altaffiltext{5}{Lawrence Berkeley National Laboratory, 1 Cyclotron Rd.,
	Berkeley, CA 94720, USA}

\altaffiltext{6}{Harvard-Smithsonian Center for Astrophysics,
	Cambridge, MA 02138, USA}

\altaffiltext{7}{LPNHE, CNRS/IN2P3, Universit\'{e} Pierre et Marie Curie Paris 6,
	Universit\'{e} Denis Diderot Paris 7,
	4 place Jussieu, 75252 Paris, France}
	
\altaffiltext{8}{Department of Physics and Astronomy, University of Wyoming,
	Laramie, WY 82071, USA}
	
\altaffiltext{9}{Department of Physics and Astronomy and PITT PACC, University of Pittsburgh,
	Pittsburgh, PA 15260, USA}

\altaffiltext{10}{Observat\'orio Nacional, Rua Gal. Jos\'e Cristino 77,
	Rio de Janeiro, RJ - 20921-400, Brazil}

\altaffiltext{11}{Laborat\'orio Interinstitucional de e-Astronomia, - LIneA,
	Rua Gal. Jos\'e Cristino 77, Rio de Janeiro, RJ - 20921-400, Brazil}

\altaffiltext{12}{Department of Physics and Astronomy, Ohio University,
	251B Clippinger Labs, Athens, OH 45701, USA}


\altaffiltext{13}{School of Physics and Astronomy, University of St Andrews,
	St Andrews, KY16 9SS, UK}

\altaffiltext{14}{Department of Physics \& Astronomy, Johns Hopkins University,
	3400 N. Charles Street, Baltimore, MD 21218, USA}
	
\begin{abstract}

We describe the \texttt{redmonster} automated redshift measurement and spectral classification
software designed for the extended Baryon Oscillation Spectroscopic
Survey (eBOSS) of the Sloan Digital Sky Survey IV (SDSS-IV).  We describe 
the algorithms, the template standard and requirements, and the newly developed
galaxy templates to be used on eBOSS spectra.  We present results from testing on early data
from eBOSS, where
we have found a 90.5\% automated redshift and spectral classification success rate for the
luminous red galaxy sample (redshifts 0.6~$\lesssim$~$z$~$\lesssim$~1.0).
The \texttt{redmonster} performance meets the eBOSS cosmology requirements for redshift classification
and catastrophic failures, and represents a significant improvement over the previous pipeline.
We describe the empirical
processes used to determine the optimum number of additive polynomial terms in our models
and an acceptable $\Delta\chi_r^2$ threshold for declaring
statistical confidence. Statistical errors on redshift measurement due to photon
shot noise are assessed, and we find typical values of a few tens of km~s$^{-1}$.
An investigation of redshift differences in repeat observations scaled by error estimates
yields a distribution with a Gaussian
mean and standard deviation of $\mu\sim$~0.01 and
$\sigma\sim$~0.65, respectively, suggesting the reported statistical redshift uncertainties
are over-estimated by $\sim$~54\%.
We assess the effects of
object magnitude, signal-to-noise ratio, fiber number, and fiber head location
on the pipeline's redshift success rate.  Finally,
we describe directions of ongoing development.

\end{abstract}


\section{Introduction}

\setcounter{footnote}{0}

Redshift surveys are a fundamental tool in modern observational astronomy. These surveys
aim to measure redshifts of galaxies, galaxy clusters, and quasars to map the 3-dimensional
distribution of matter.  These observations allow measurements
of the statistical properties of the large-scale structure of the universe.  In conjunction
with observations of the cosmic microwave background, redshift surveys can also be used
to place constraints on cosmological parameters, such as the Hubble constant
(e.g., \citealp{beu2011})
and the dark energy equation of state through measurements of the baryon acoustic
oscillation (BAO) peak, first detected in the clustering of galaxies (\citealp{eis2005}, \citealp{col2005}).
The first systematic redshift survey was the CfA Redshift Survey \citep{dav1982}, 
measuring redshifts for approximately 2,200 galaxies.  Such early surveys were limited in scale due
to single object spectroscopy.  The development of fiber-optic and multi-slit
spectrographs enabled the simultaneous observations of hundreds or thousands of spectra,
making possible much larger surveys, such as the DEEP2 Redshift Survey \citep{new2013},
the 6dF Galaxy Survey (6dFGS; \citealp{jon2004}), Galaxy and Mass Assembly (GAMA; \citealp{lis2015}),
and the VIMOS Public Extragalactic Survey (VIPERS; \citealp{gar2014}),
measuring redshifts for approximately
50,000, 136,000, 300,000, and 55,000 objects, respectively.

The Sloan Digital Sky Survey (SDSS; \citealp{yor2000}) is the largest redshift survey
undertaken to date.  At the conclusion of SDSS-III, the third iteration of SDSS
\citep{eis2011}, a total of 4,355,200 spectra had been obtained.  Of these, 2,497,484 were
taken as part of the Baryon Oscillation Spectroscopic Survey (BOSS; \citealp{daw2013}),
containing 1,480,945 galaxies, 350,793 quasars, and 274,811 stars.
The ``constant mass'' (CMASS) subset of the BOSS sample is composed
of massive galaxies over the approximate
redshift range of $0.4 < z < 0.8$ and typical S/N values of $\sim 5$/pixel.  The automated redshift
measurement and spectral classification of such large numbers of objects presents a
challenge, inspiring refinements of the \texttt{spectro1d} pipeline \citep{bol2012}.
This software models each co-added spectrum as a linear combination of
principal component analysis (PCA) basis
vectors and polynomial nuisance vectors and adopts the
combination that produces the minimum $\chi^2$ as the
output classification and redshift.  PCA-reconstructed models were chosen
due to their close ties to the data, allowing the PCA eigenspectra to potentially capture any intrinsic populations
within the training sample. This pipeline was able to achieve an automated
classification success rate of 98.7\% on the CMASS sample (and 99.9\% on the
lower-redshift, higher-S/N LOWZ sample).  However, the software was only able to
successfully classify 79\% of the quasar sample, 
which resulted in the need for the entire sample to be visually inspected \citep{par2012}.

The Sloan Digital Sky Survey IV (SDSS-IV; \citealp{bla2016}) is the fourth iteration of the SDSS.
Within SDSS-IV, the Extended Baryon Oscillation Spectroscopic Survey (eBOSS; \citealp{daw2016})
will precisely measure the expansion history of the Universe throughout eighty percent
of cosmic time through observations of galaxies and quasars in a range of redshifts
left unexplored by previous redshift surveys.  Ultimately, eBOSS
plans to use approximately 300,000 luminous red galaxies (LRGs; 0.6~$<$~$z$~$<$~1.0),
200,000 emission line galaxies (ELGs; 0.7~$<$~$z$~$<$~1.1), and 
700,000 quasars (0.9~$<$~$z$~$<$~3.5) to measure the clustering of matter.

The primary science goal of eBOSS is to measure the length scale of the BAO
feature in the spatial correlation function
in four discrete redshift intervals to $1-2$\% precision,
thereby constraining the nature of the dark energy that drives the accelerated expansion
of the present-day universe.
A set of requirements for the redshift and classification pipeline
was established to meet these goals.  As given in \citet{daw2016}, these
requirements are: (1) redshift accuracy $c\sigma_z/(1+z)<$~300 km~s$^{-1}$ for
all tracers at redshift $z<1.5$ and
$<(300+400(z-1.5))$ km~s$^{-1}$ RMS for quasars at $z>1.5$ ; and (2) fewer than 1\%
unrecognized redshift errors of $>1000$ km~s$^{-1}$ for LRGs and $>3000$ km~s$^{-1}$ for quasars
(referred to in this paper as ``catastrophic redshift failures").  Additionally, the pipeline should
return confident redshift measurements and classifications for $>90$\% of spectra.

The higher-redshift, lower-S/N (typically $\sim 2$/pixel for galaxies) targets in eBOSS present a
new challenge for automated redshift measurement and classification software.  Initial tests
with the \texttt{spectro1d} PCA basis vectors
predicted success rates of $\sim 70$\% for the LRG sample, which is well below the
specified science requirements.  This is due, in part, to the flexibility in fitting PCA
components to a spectrum, which allows non-physical combinations of basis vectors to pollute
the redshift measurements and statistical confidences thereof.  Additionally, while possible (e.g.,
\citealp{che2012}), mapping PCA coefficients onto physical properties is a difficult task.  It requires
the use of a transformation matrix, and confidence in the results is, at best, unintuitive, and possibly
uncertain.

To meet these challenges, we have developed an archetype-based software system
for redshift measurement and spectral classification named \texttt{redmonster}.  We have developed
a set of theoretical templates from which spectra can be classified.  \texttt{redmonster}
is written in the Python programming language. The project is open source,
and is maintained on the first
author's GitHub account\footnote{https://github.com/timahutchinson/redmonster}.
The analysis performed in this paper uses tagged version \texttt{v1\_0\_0}.  The development
of this software was driven by the following goals:

\begin{enumerate}
\item Redshift measurement and classification on the basis of
discrete, non-negative, and physically-motivated model spectra;
\item Robustness against unphysical PCA solutions likely to arise for low-S/N ELG and
LRG spectra in eBOSS, particularly in the presence of imperfect sky-subtraction;
\item Determination of joint likelihood functions over redshift and physical parameters;
\item Self-consistent determination and application of hierarchical redshift priors;
\item Self-consistent incorporation of photometry and spectroscopy in performing redshift constraints;
\item Simultaneous redshift and parameter fits to each individual exposure in multi-exposure data;
\item Custom configurability of spectroscopic templates for different target classes;
\item Automated identification of multi-object superposition spectra.
\end{enumerate}

The software as described in this paper meets design goals 1, 2, 3, and 7.  We have chosen
to enumerate the full list of design goals to provide a forward-looking vision of new and interesting
possibilities.

In this paper, we describe \texttt{redmonster} and its application to the eBOSS LRG sample.
The organization is as follows:  Section~\ref{sec:software} describes
automated redshift and classification algorithms and procedures of \texttt{redmonster}.
We include the requirements and standardized format for templates and a description
of the eBOSS galaxy and star templates in Section~\ref{sec:templates}.  The
core redshift measurement algorithm is described in Section~\ref{sec:fitting} and Section~\ref{sec:interpretation}.
Section~\ref{sec:parameters}
gives an overview of the spectroscopic data sample of eBOSS and 
an analysis of the tuning and performance of the software on eBOSS data, including 
completeness and purity. Section~\ref{sec:classification} provides a description
of the classification of eBOSS LRG spectra, including redshift success dependence,
effects on the final redshift distribution in eBOSS, and precision and accuracy.
Finally, Section~\ref{sec:conclusion} provides
a summary and conclusion.  The content and structure of the output files of \texttt{redmonster}
are described in Appendix~\ref{sec:output}.

\section{Software Overview} \label{sec:software}

The use of physically-motivated templates allows mapping of physical properties
from the best-fitting template onto the model used to create the suite of templates.
To better facilitate the exploration of large, multi-dimensional
parameter spaces, spectral fitting is performed in Fourier space, allowing \texttt{redmonster}
to combine the speed of cross-correlation techniques with the statistical framework of
forward-modeling.  Additionally, significant effort has been made to write software not specific
solely to SDSS, but rather in a manner that facilitates use in other redshift surveys, such as the
Dark Energy Spectroscopic Instrument (DESI; \citealp{lev2013}).
We have also prioritized end-user customizability in the way the software operates.

Informed by our design goal of developing survey-agnostic software,
the current version of \texttt{redmonster}
requires only the following data as input:
\begin{enumerate}
\item Wavelength-calibrated, sky-subtracted, flux-calibrated, and co-added spectra,
rebinned onto a uniform baseline of constant $\Delta$log$_{10}\lambda$ per pixel;
\item Statistical error-estimate vectors for each spectrum, expressed as inverse variance.
\end{enumerate}

While SDSS spectra are shifted such that measured velocities will be relative to the solar system
barycenter at the mid-point of each 15-minute exposure, no such requirement exists for
\texttt{redmonster} input spectra.  Redshifts will be measured relative to a frame of the end-user's choosing.

\subsection{Templates} \label{sec:templates}

In order to make redshift and parameter measurements and select among galaxy, quasar,
and stellar (and possibly other) object types with the highest statistical confidence,
the pipeline requires a set of
templates that both spans the entire space of object types within the survey and covers the full
wavelength range of the spectrograph over the redshift range of interest.  To this end,
\texttt{redmonster} uses ``archetype'' template grids, where ``archetype" refers to
single spectral templates that are not fit in linear combination with any other
templates, excluding low-order polynomial nuisance vectors.
A series of archetype spectral templates spanning the relevant parameter space is recorded
in an \texttt{ndArch.fits} (signifying N-dimensional archetypes) file.
These \texttt{ndArch.fits}
files conform to a standard that requires a general form of template spectra
written by end users and ingested into the redshift software.  The format allows
highly configurable spectroscopic template classes without any re-coding of the
low-level fitting routines. This file standard is oriented towards the familiar units
and conventions of optical spectroscopic redshift measurement.
Reader and writer routines for \texttt{ndArch.fits} files conforming
to this standard are included in the \texttt{redmonster}
package. A brief description of the standard is given here, while full documentation
can be found in the software package.

The data contained in an \texttt{ndArch.fits} file
consists of a single multi-dimensional array
that contains all possible spectral templates for a class of object,
containing flux densities or luminosity densities in units of
$F_{\lambda}$ (power per unit area per unit wavelength) or $L_{\lambda}$
(power per unit wavelength).
The absolute normalization may be physically
meaningful, but is not required to be so.
The first axis of the data array corresponds to vacuum
wavelength and is gridded in positive increments of constant
$\log\lambda$.

There may be one or more axes in addition
to the first axis, up to the maximum number allowed by
the FITS standard.  Each axis beyond the first will generally correspond
to a monotonically ordered
physical model-parameter dimension (age, metallicity,
emission-line strength, etc.), but may also correspond to
an arbitrary labeled or unlabeled collection.  The archetype template vectors are assumed to have
a uniform resolution characterized by a Gaussian
line-spread function with a dispersion parameter
$\sigma$ equal to one sampling pixel.

Templates are segregated into template classes, with each class corresponding to a single
object type of interest and contained in a single \texttt{ndArch.fits} file.  These classes are used
by \texttt{redmonster} to classify the object type of a given spectrum.
Three template classes have been developed for use in the eBOSS pipeline.  Galaxy templates
for LRG targets are described in \S\ref{sec:lrgs}.
Quasar and stellar templates have been developed and are included with \texttt{redmonster}.
Because this paper focuses primarily on performance of \texttt{redmonster} spectroscopic classifications
on the eBOSS LRG target sample, these
templates function primarily to identify non-galaxy objects that arise due to targeting impurities.
As such, we defer description of quasar and stellar templates to a later work.

\subsubsection{Galaxy templates} \label{sec:lrgs}

\begin{figure*}
\plotone{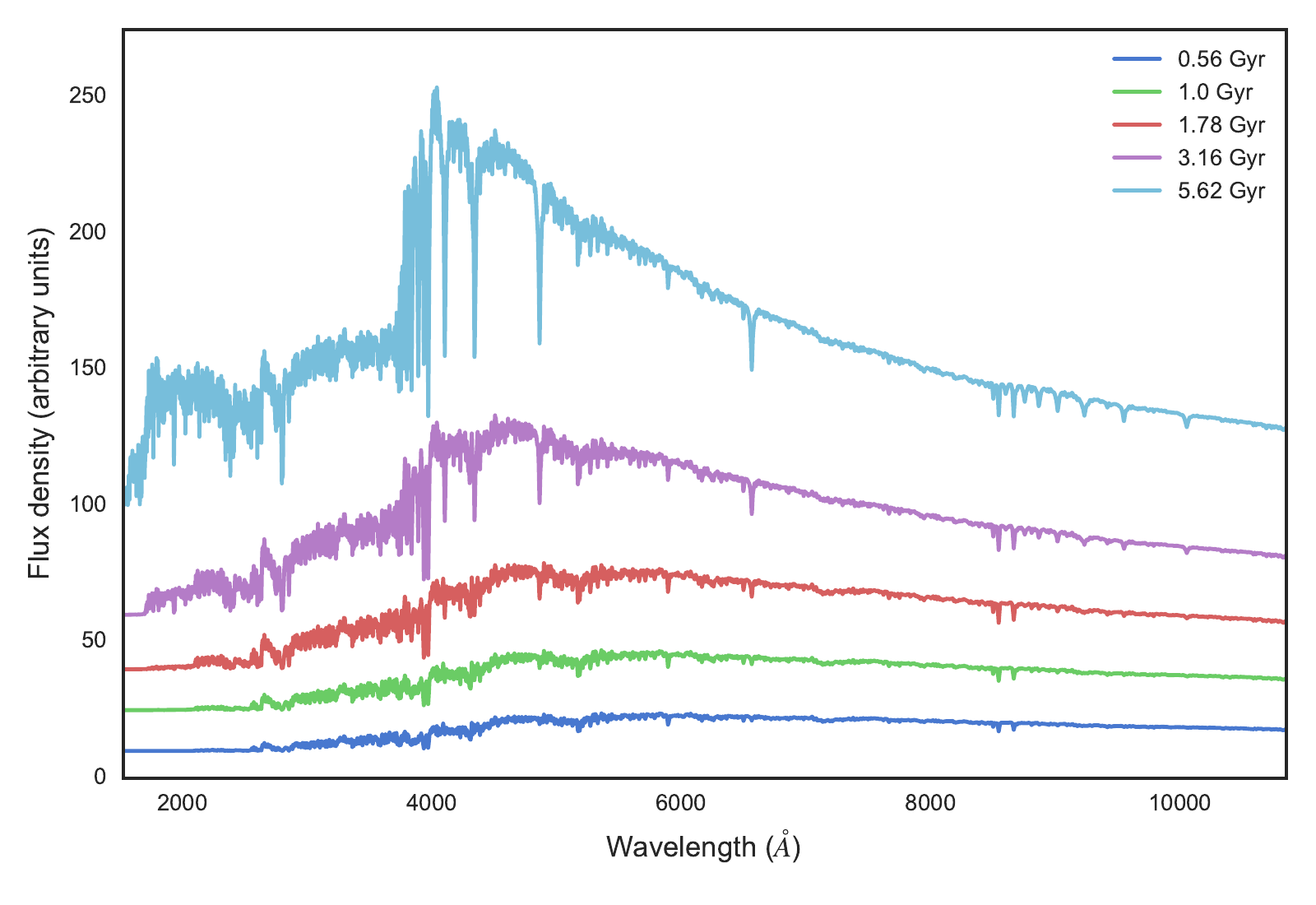}
\caption{Example LRG templates of various ages.  The templates have been vertically offset
for increased visibility.\\
\label{fig:templates}}
\end{figure*}

Our LRG templates are selected from the Flexible Stellar Population Synthesis (FSPS) model
suite (\citealp{con2009}, \citealp{con2010}) with the Padova isochrones \citep{mar2008}
and a Kroupa IMF \citep{kro2001}.  The spectra used in the models are custom high resolution theoretical
spectra \citep{con2016}; the resulting models are referred to as
FSPS-C3K.  The synthetic spectral library was constructed with the latest set of atomic and
molecular line lists (courtesy of R. Kurucz) and is based on the Kurucz suite of spectral
synthesis and stellar atmosphere routines
(SYNTHE and ATLAS12; \citealp{kur1970}, \citealp{kur1993}; \citealp{kur1981}).
The grid of spectra was computed assuming the
\citet{asp2009} solar abundance pattern and a constant microturbulence of 2 km~s$^{-1}$.
For further details see \citet{con2012}.  Table~\ref{tab:galparams} lists the physical
parameters of the LRG template suite, their range, and the frequency with which they
are binned. All models are solar metallicity.
Several example templates are shown in Figure~\ref{fig:templates}.  The templates
have been vertically offset by 800, 600, 400, 250, and 100 for visual clarity.
The wavelength range of the BOSS spectrograph extends from $3600$ \AA\ to
$1.04\ \mu$m, while the templates span the range $1525$ \AA\ $<\lambda<10852$ \AA,
allowing the templates to span the redshift range $0<z<1.36$.  These templates will
be extended further into the blue to cover the full redshift range of the ELG sample in SDSS DR14.

\begin{deluxetable}{cll}
\tabletypesize{\footnotesize}
\tablecolumns{3}
\tablewidth{0pt}
\tablecaption{\label{tab:galparams} Galaxy template physical parameter dimensions and sampling}
\tablehead{
\colhead{Parameter} & \colhead{Range} & \colhead{$N_{samples}$}
}
\startdata
$\log_{10}\mathrm{(Age)}$ & [-2.5, 1] Gyr & 15\\
$\log_{10}\sigma_{vel}$ & [2, 2.6] km~s$^{-1}$ & 4\\
H$\alpha$ EW & [0, 100] \AA & 5
\enddata
\end{deluxetable}

\subsubsection{Stellar templates} \label{sec:stars}

Our stellar templates are synthetic spectra computed using Kurucz ATLAS9
models by \citet{mes2012} and the radiative transfer code ASS$\epsilon$T
\citep{koe2009}. The reference solar abundances are from \citet{asp2005},
with enhanced $\alpha$-element abundances, mimicking the trends found in the
Milky Way, and a constant micro-turbulence velocity of 2 km s$^{-1}$.
Continuum opacity is based on the Opacity Project and Iron Project
photoionization cross-sections, collected by \citet{all2003}
and \citet{all2008}, and line opacities are based on data compiled
by Kurucz\footnote{kurucz.harvard.edu}, with updates from \citet{bar2000}.
Line absorption coefficients for Balmer lines are computed with the codes provided
by \citet{bar2015}.

\subsection{Spectral fitting} \label{sec:fitting}

The \texttt{redmonster} software approaches redshift measurement and classification as a $\chi^2$ 
minimization problem by cross-correlating the observed spectrum with each spectral template
in the \texttt{ndArch.fits} file over a discretely sampled redshift interval.
The algorithm is similar to the method described by \citet{ton1979}.
By default, pixels with S/N $> 200$ (which likely indicates a cosmic ray)
or with $f_\lambda < -10\sigma$ (unphysical negative flux at $10\sigma$ significance),
are masked for each observed spectrum.
These values are configurable when running the software.
The spectrum is then fit with an error-weighted least-squares linear 
combination of a single template and a low-order polynomial across the range of trial redshifts.
The polynomial term serves to absorb Galactic and intrinsic interstellar extinction,
sky-subtraction residuals, and spectro-photometric
calibration errors not accounted in the templates. By default, trial redshifts are separated by
the spacing set by a single pixel in wavelength, though this, too, is configurable.
For the eBOSS classifications, we have chosen to separate trial redshifts by
one pixel for stars, two pixels for galaxies, and four pixels for quasars ($\sim69$ km~s$^{-1}$,
$\sim138$ km~s$^{-1}$, and $\sim278$ km~s$^{-1}$, respectively) to reduce computation time.
A redshift range must also be specified for each \texttt{ndArch.fits} file being used; for the eBOSS
data, we use $-0.1<z<1.2$ for galaxies, $-0.005<z<0.005$ for stars, and $0.4<z<3.5$ for quasars.
This fitting process results in a $\chi^2$ value for a spectral template at a single trial redshift.
Repeating this fit for all templates within the \texttt{ndArch.fits} and across all trial redshifts defines
a $\chi^2(\vec{P},z)$ surface, where $\vec{P}$ is a vector spanning the parameter-space
of the template class.

The software is able to explore large parameter spaces within a template class by
pre-computing some matrix elements in Fourier space during
the fitting process (see also \citealp{gla1998}).  In order to fit a set of data $\vec{d}$ with a set of $n$ basis vectors
$\{\vec{x}_k\}$ in a minimum $\chi^2$ sense, the model $\vec{m}$ takes the form
\begin{equation}
\vec{m}=\sum\limits_{k=1}^n a_k \vec{x}_k,
\end{equation}
where, in the case of \texttt{redmonster}, $\{\vec{x}_k\}$ consists of a single physical
template and $n-1$ polynomial terms.
Thus, the $\chi^2$ of the model relative to the data, assuming $N$ data points, is
\begin{equation}
\chi^2 = \sum\limits_{i=1}^N \frac{(d_i - \sum\limits_{k=1}^n a_k x_{k,i})^2}{\sigma_i^2}
\end{equation}
where $\sigma_i^2$ is the statistical variance of the $i^{\mathrm{th}}$ pixel of $\vec{d}$.  We find
the $j^{\mathrm{th}}$ maximum likelihood estimator, $\hat{a}_j$, by minimizing $\chi^2(\vec{a})$
with respect to $a_j$.  Here, $a_j$ is an arbitrarily chosen basis vector coefficient from $\vec{a}$,
as the result is independent of our choice.  Solving
$\partial{\chi^2}/\partial{a_j}~=~0$
yields
\begin{equation} \label{eq:4}
\sum\limits_{i=1}^N\sum\limits_{k=1}^n \frac{\hat{a}_k x_{k,i} x_{j,i}}{\sigma_i^2} = 
\sum\limits_{i=1}^N \frac{d_i x_{j,i}}{\sigma_i^2}.
\end{equation}
Because eBOSS spectra are binned linearly in $\mathbf{\log_{10}}\lambda$, velocity redshifts are uniform
linear shifts in pixel space.  Thus, fitting at a trial redshift introduces a ``lag" in pixel-space $l$
of the basis vectors relative to the data, after which estimators become
a function of $l$, and Eq.~\ref{eq:4} becomes
\begin{equation} \label{eq:5}
\sum\limits_{i=1}^N\sum\limits_{k=1}^n \frac{\hat{a}_k(l) x_{k,i-l} x_{j,i-l}}{\sigma_i^2} = 
\sum\limits_{i=1}^N \frac{d_i x_{j,i-l}}{\sigma_i^2}.
\end{equation}
This result can be cast in terms of matrices:
\begin{equation}
(\textbf{P}^\top \textbf{N}^{-1} \textbf{P}) \vec{\hat{a}} = (\textbf{P}^\top \textbf{N}^{-1}) \vec{d}
\end{equation}
where $\textbf{P}$ is an $(N \times n)$ matrix with each column a basis vector,
$\textbf{N}$ is an $(N \times N)$ diagonal matrix with $\textbf{N}_{ii} =
$ $\sigma_i^2$, and $\vec{\hat{a}}$ is the vector of maximum likelihood esimators
for which we wish to solve.  The product $\textbf{P}^\top \textbf{N}^{-1} \textbf{P}$
is the correlation matrix of the templates, with elements
\begin{equation}
(\textbf{P}^\top \textbf{N}^{-1} \textbf{P})_{jk} = \sum\limits_{i=1}^N \frac{x_{j,i-l} x_{k,i-l}}{\sigma_i^2},
\end{equation}
while the elements $(\textbf{P}^\top \textbf{N}^{-1}\vec{d})_{jk}$ are given by the
right-hand side of Eq.~\ref{eq:5}.
Over a range of continuous $l$, these elements are the discrete convolutions
$(1/\vec{\sigma^2})*(\vec{x}_j\cdot\vec{x}_k)(l)$ and $(\vec{d}/\vec{\sigma^2})*\vec{x_j}(l)$,
respectively, and may be computed as products
in Fourier space.  Since the polynomial terms have no physical meaning,
they may remain fixed in the observed frame of the data (i.e., have no $l$ dependence),
and it is possible to pre-compute the fraction $(n-1)^2/n^2$ of the total matrix elements for a
template class, greatly reducing computational requirements.

One of the motivations for the development of \texttt{redmonster} and its galaxy templates
was the restriction that all redshifts and classifications must be derived from physical models.
To that end, the resulting $\hat{a}_k(l)$ at each point in redshift-parameter space
is evaluated for physicality based on the coefficient of the template component of the model.
Fits with a negative template amplitude are rejected, and that point in the $\chi^2(\vec{P},z)$
surface is assigned a value of $\chi_{\mathrm{null}}^2$,
where $\chi_{\mathrm{null}}^2$ is defined as the $\chi^2$ of the best-fit model
where the amplitude of the template coefficient is forced to 0 (i.e., a polynomial-only model).
Because the best-fit models produce the lowest possible $\chi^2$ value at each point in
redshift-parameter space, and because the value of $\chi_{\mathrm{null}}^2$ is a function only of
the data itself, the effect of this physicality constraint is to introduce a flat ``ceiling"
in the $\chi^2(\vec{P},z)$ surface, while maintaining continuity.

\begin{deluxetable*}{cll}
\tablecolumns{3}
\tablewidth{0pt}
\tablecaption{\label{tab:zwarning} \texttt{redmonster} ZWARNING bit-mask definitions}
\tablehead{
\colhead{Bit} & \colhead{Name} & \colhead{Definition}
}
\startdata
0 & \texttt{SKY} & Sky fiber\\
1 & \texttt{LITTLE\_COVERAGE} & Insufficient wavelength coverage\\
2 & \texttt{SMALL\_DELTA\_CHI2} & $\chi^2$ of best fit is too close to that of second best
($<0.01$ in $\chi_r^2$)\\
5 & \texttt{Z\_FITLIMIT} & $\chi_{\mathrm{min}}^2$ at edge of the redshift fitting range (\texttt{Z\_ERR}
set to -1)\\
7 & \texttt{UNPLUGGED} & Fiber was broken or unplugged, and no spectrum was obtained\\
8 & \texttt{NULL\_FIT} & At least one template class had constant-valued $\chi^2$ surface
\enddata
\end{deluxetable*}

\subsection{$\chi^2$ interpretation} \label{sec:interpretation}

After the computation of the $\chi^2(\vec{P},z)$ surface for each template class,
the minimum $\chi^2$ in spanning all templates is found at each trial redshift. The resulting series
of best fits defines
a $\chi^2(z)$ curve for each template class (similar to Figure 2 of \citealp{bol2012}).
Interpolation over this curve is then performed using a cubic B-spline
(i.e., a C2-continuous composite B\'ezier curve), through which all local minima are identified.
The $N$ best redshifts for a particular template class are defined by the $N$ lowest minima of the
$\chi^2(z)$ curve.  The curve around each of these
minima is then fit by a quadratic function using the three points nearest the minimum.
The analytic minima of each quadratic fit are adopted as the spectrum's candidate redshifts.  The
statistical error on each candidate is evaluated as the change in redshift $\pm\delta z$
for which the $\chi^2$ of the quadratic fit increases by one from its minimum value.

In the event the global minimum
falls on the edge of the explored redshift range, the \texttt{Z\_FITLIMIT} bit
is triggered within the \texttt{ZWARNING} bit-mask,
and both \texttt{Z} and \texttt{Z\_ERR} are set to -1.  The definitions of all failure modes
captured by \texttt{ZWARNING} are shown in Table~\ref{tab:zwarning}.
Bit-masks 0-7 are identical to those of \texttt{spectro1d},
although bits 3, 4 and 6 are not used by \texttt{redmonster} and are retained only for consistency.
Bit 4 (\texttt{MANY\_OUTLIERS}) was found in SDSS-III to flag too many good quasar redshifts.
Bit 6 (\texttt{NEGATIVE\_EMISSION}) has been deprecated, allowing all spectra to be considered.
The \texttt{NEGATIVE\_MODEL} bit (bit 3) is unnecessary as \texttt{redmonster} restricts the
template amplitude to be positive for reasons of physicality.
Bit 8 is new and is triggered in the rare case of a template
class having a $\chi^2$ surface with no local minima.
The output files are discussed in detail in Appendix~\ref{sec:output}.

\begin{figure*}
\plotone{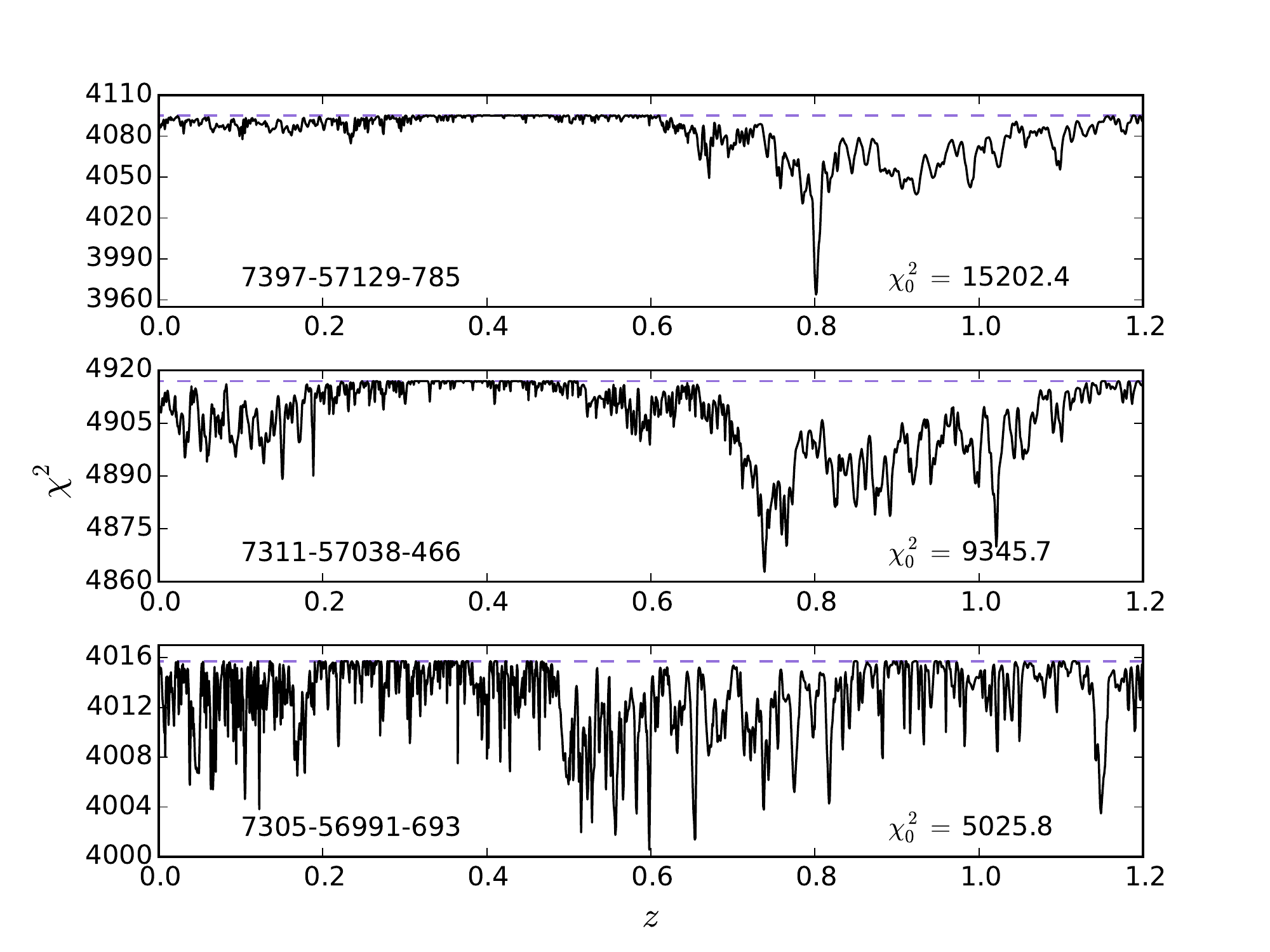}
\caption{Example $\chi^2(z)$ curves from the fits to spectra of three different LRG targets.
The spectra are labeled by \texttt{PLATE}-\texttt{MJD}-\texttt{FIBERID}.  The dashed purple
lines show the value of $\chi_\mathrm{null}^2$.
The $\chi_0^2$ values shown are the $\chi^2$ of a $\vec{0}$ model (i.e., the (S/N)$^2$ of the data).
Note that the vertical axis shows raw $\chi^2$, and must be scaled by the number of
degrees of freedom (i.e., the number of unmasked spectrum pixels minus the number of components
in the fit) for comparison with $\Delta\chi_\mathrm{threshold}^2$.
The top panel shows a $z=0.802$ galaxy with \texttt{ZWARNING} $=0$,
indicating a confident redshift measurement and classification.  The middle panel shows an
LRG target with statistically indistinguishable fits at $z=0.738$ and $z=1.021$, triggering
the \texttt{SMALL\_DELTA\_CHI2} flag (\texttt{ZWARNING} $=4$).  The bottom panel shows
an LRG target with no $\chi^2$ minima separated from $\chi_{\mathrm{null}}^2$ with
significance greater than $\Delta\chi_\mathrm{threshold}^2=0.005$.
\label{fig:chi2curves}}
\end{figure*}

Due to noise, some spectra can have multiple minima
in the vicinity of one another that are not statistically significant.  For all template
classes, we ignore local minima that are separated in redshift from a lower-$\chi^2$ minimum by 
less than a given threshold, $\Delta$V.  Local minima separated by more than $\Delta$V
are explicitly evaluated, since they constitute redshift failures if they are 
statistically indistinguishable from one another.
We define a quantity, $\Delta \chi_\mathrm{threshold}^2$, as the minimum acceptable
 $\Delta\chi_\mathrm{red}^2=(\chi_2^2-\chi_1^2)/$dof, where $\chi_1^2$ corresponds to
the global minimum and $\chi_2^2$ corresponds to some secondary minimum.
The $\Delta \chi_\mathrm{red}^2$ between two minima must be greater than this threshold
for statistical confidence to be declared.
Values of $\Delta \chi_\mathrm{red}^2$ less than
this threshold will trigger the \texttt{SMALL\_DELTA\_CHI2} bit in the \texttt{ZWARNING} bit-mask,
indicating a lack of statistical confidence.
This process is illustrated schematically in Figure 2 of \citet{bol2012}; example
$\chi^2(z)$ curves from eBOSS spectra are shown in Figure~\ref{fig:chi2curves}.

The fits are then compared across template classes if multiple \texttt{ndArch} files were used.
The $N$ best fits from each template class are combined into a single set, and are then
sorted in ascending order of $\chi_\mathrm{{red}}^2$.
The $N$ best fits from this combined set are adopted as the
best redshifts and classifications for the object.  These fits are re-evaluated for statistical
confidence and flagged according to the above criteria, although $\chi_\mathrm{min}^2$ separated
by less than $\Delta$V will no longer trigger the \texttt{SMALL\_DELTA\_CHI2} flag.  Additionally,
redshift differences between two classes less than the quadrature sum of the error estimates
are not flagged even if the $\Delta \chi^2$ is below the threshold, as the redshift, and not the class,
is the primary measurement being made.  By default, the software keeps five redshifts and
classifications per spectrum ($N=5$), but this is user-configurable.

We note here that, strictly speaking, the use of minimium-$\chi^2$ regression should be
limited to cases where the measurement errors account for all of the statistical
scatter and the model is correctly specified (i.e., the template class is correct).  Otherwise,
the parameter values and their confidence intervals may not be scientifically meaningful.
While the measurement errors of eBOSS data do not account for all of the scatter and the
chosen template family is not a complete representation of the galaxies present in the sample,
we empirically calibrate the robustness of our statistics through the use of repeat observations,
as discussed in \S\ref{sec:precision}.

\section{Optimization of \texttt{redmonster} parameters for
\MakeLowercase{e}BOSS LRGs} \label{sec:parameters}

\begin{figure*}
\includegraphics[width=\textwidth]{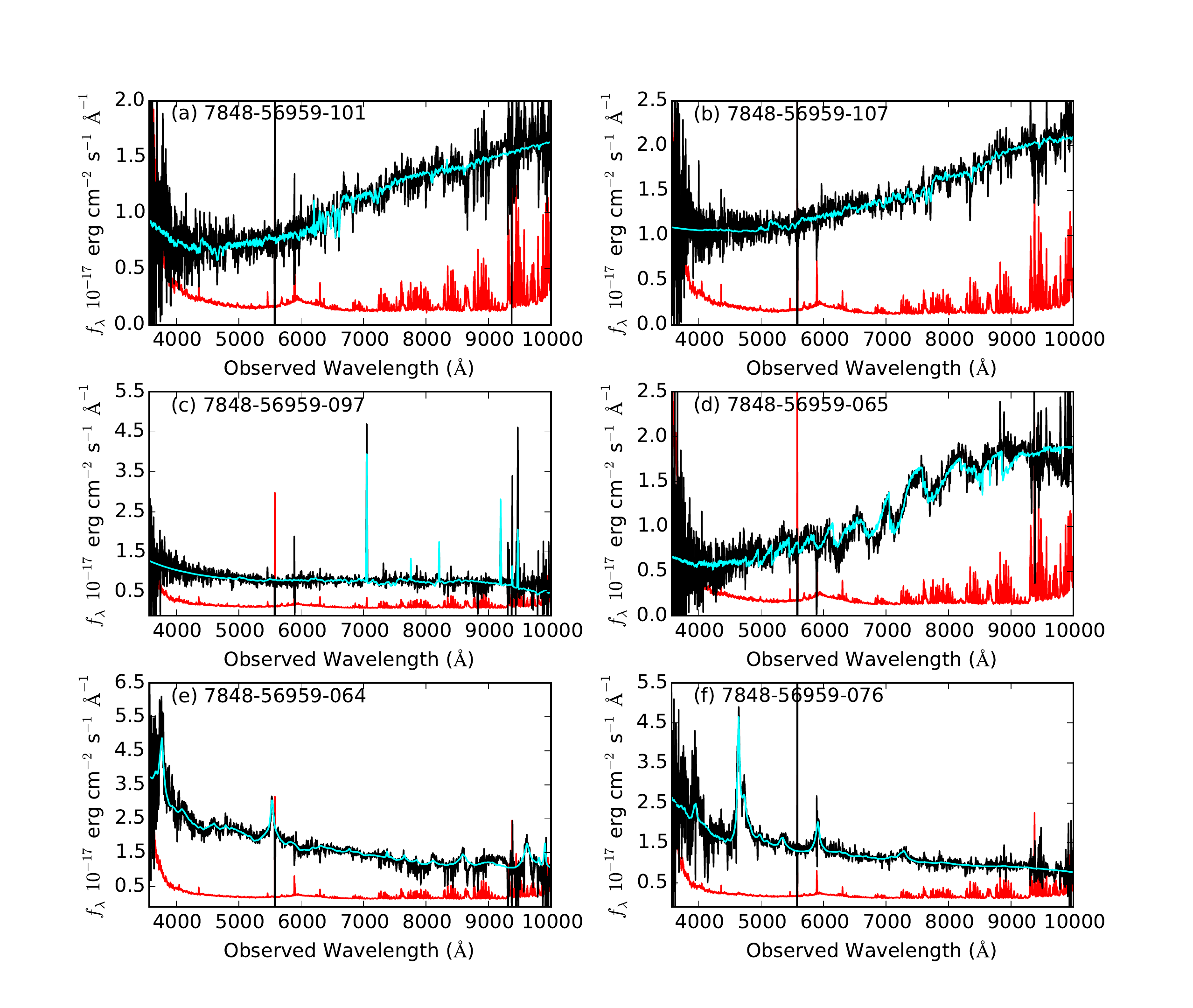}
\caption{Example eBOSS spectra.  The data, smoothed over a 5-pixel window,
are presented in black, the best-fit model from \texttt{redmonster} is shown in cyan, and the 1-$\sigma$ error
(as estimated by \texttt{idlspec2d}) is shown in red.  Each spectrum has been labeled
with its unique \texttt{PLATE-MJD-FIBERID}.  The objects shown here are: (a) LRG-target
galaxy, $z=0.664$; (b) LRG-target galaxy, $z=0.943$; (c) ELG-target galaxy, 
$z=0.891$; (d) M-star; (e) quasar, $z=0.978$; (f) quasar, $z=2.823$.\\
\label{fig:6panel}}
\end{figure*}

The main targets for eBOSS spectroscopy consist of LRGs at redshifts $0.6<z<1.0$
\citep{pra2015}, ELGs in the
range $0.6<z<1.1$ (\citealp{com2015}, \citealp{rai2016}, \citealp{del2016}),
``clustering" quasars in the range $0.9<z<2.2$,
re-observations of faint Ly$\alpha$ quasars in the range $2.1<z<3.5$, and new
Ly$\alpha$ quasars at redshifts $2.1<z<3.5$ \citep{mye2015}.
A selection of example eBOSS spectra is shown in Figure~\ref{fig:6panel}.  The 
spectral classification and redshift software described here will be applied to all spectra
obtained with the BOSS spectrograph, including targets outside the LRG, ELG, and quasar
selection algorithms.  Here we demonstrate the tuning of \texttt{redmonster} parameters
and the performance of \texttt{redmonster} on the LRG target sample from eBOSS.

The LRG target class consists of massive red galaxies that have been color-selected using
SDSS imaging \citep{gun1998} in \textit{ugriz} filters \citep{fuk1996} and imaging from the
Wide-field Infrared Survey Explorer (WISE; \citealp{wri2010}).  These targets have a faint
magnitude limit of $i<21.8$ (AB).  A full investigation of the LRG
selection is presented in \citet{pra2015}.

Spectroscopic data for eBOSS are obtained using the 2.5-m Sloan Telescope
at Apache Point Observatory \citep{gun2006} with the BOSS spectrograph
system.  The BOSS instrument is composed of two double-arm spectrographs
fed by 1000 optical fibers plugged
into a drilled aluminum plate that is positioned in the telescope focal plane.
A summary of this system is given in
Table 2 of \citet{bol2012} and a full account is given in \citet{sme2013}.
Each of the optical fibers feeding the spectrographs is numbered with a 
\texttt{FIBERID} index ranging from 1-1000.  Each physical plug plate is given
a unique \texttt{PLATE} number.  Finally,
because a given \texttt{PLATE} may be observed more than once with different 
mappings between \texttt{FIBERID} and target spectra, each plugging is given an
\texttt{MJD} number corresponding to the modified Julian date of the observation.  Thus,
any combination of \texttt{PLATE}, \texttt{MJD}, and \texttt{FIBERID} constitutes
a unique eBOSS co-added spectrum.  Pluggings are observed in 15-minute exposures, which
are co-added during the data reduction process as described in
\citet{daw2013}.  The 1D spectral outputs of this calibration, extraction, and co-addition process
are stored in the ``\texttt{spPlate}'' FITS files, which become the inputs to the 
software described in this paper.

Two significant changes have been made to the spectral extraction and co-addition of individual
exposures for the second data release (DR14) of SDSS-IV.  These changes were developed
to improve classification of lower signal-to-noise data in eBOSS.  The first major change
improves the way atmospheric differential refraction (ADR) corrections are applied
to eBOSS spectra, and is described in \citet{jen2016}.
The improved ADR corrections are similar to prior work (\citealp{mar2015}, \citealp{har2016})
and primarily improve flux calibrations for quasar spectra.
The second change corrects a known bias
in the co-addition of individual exposures.  This correction has significant impact on the classification
of galaxy spectra; thus, we provide a description below.






Individual exposures are initially flux calibrated with no constraint that the same object has the
same flux across different exposures.  Empirical ``fluxcorr'' vectors are broadband corrections to
bring the different exposures into alignment for each object prior to co-addition.  In DR13 and prior,
these were implemented for each spectrum by minimizing

\begin{equation}
    \chi^2_i = \sum_\lambda {
        (f_{i\lambda} - f_{\mathrm{ref},\lambda} / a_{i\lambda} )^2 \over
        (\sigma_{i\lambda}^2 + \sigma_{\mathrm{ref},\lambda}^2 / a_{i\lambda}^2)
        }
\end{equation}

where $f_{i\lambda}$ is the flux of exposure $i$ at wavelength $\lambda$,
$f_{\mathrm{ref},\lambda}$ is the flux of the selected reference exposure,
and $a_{i\lambda}$ are low order Legendre polynomials.
The number of polynomial terms is dynamic, up to
a maximum of 5 terms.  Higher order terms are added only if they improve
the $\chi^2$ by 5 compared to one less term.
This approach is biased toward small $a_{i\lambda}$ since that inflates the denominator to
reduce the $\chi^2$.

For DR14, we solve the fluxcorr vectors relative to a common weighted co-add $F_\lambda$
which is treated as noiseless compared to the individual exposures,

\begin{equation}
    \chi^2_i = \sum_\lambda {
        (f_{i\lambda} - F_\lambda / a_{i\lambda})^2 \over
        \sigma_{i\lambda}^2
        }.
\end{equation}

\begin{figure}
\plotone{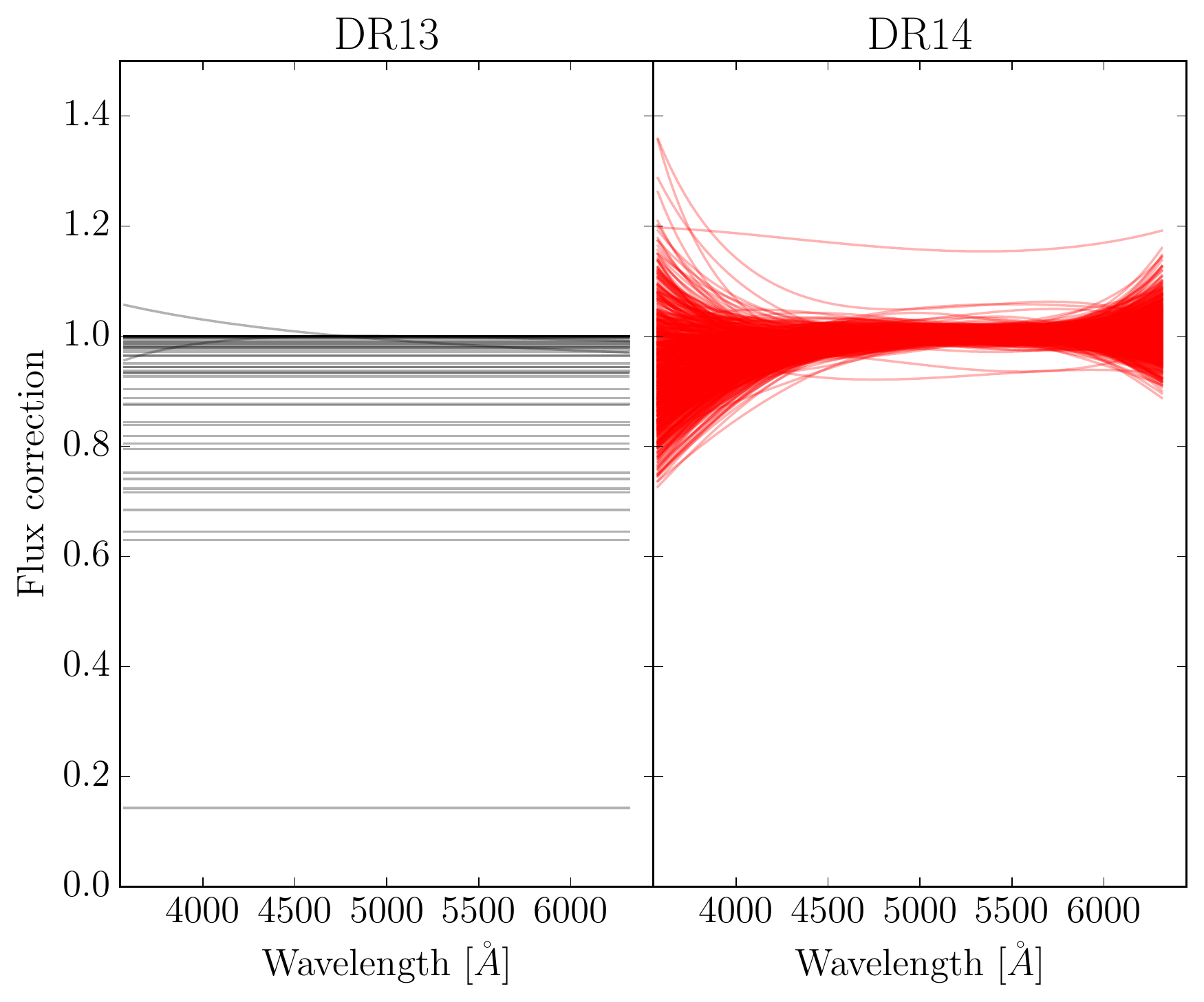}
\caption{Calibration corrections for 5 exposures of 155 LRG spectra on plate 7898 (blue camera)
in DR13 (left panel) and DR14 (right panel).\\
\label{fig:fluxcorr}}
\end{figure}

We additionally include an empirically tuned prior that $a_{i\lambda} \sim 1$ to avoid large
excursions in the solution for very low signal-to-noise data.
Figure~\ref{fig:fluxcorr} shows the fluxcorr corrections for five exposures of 155 LRGs. 
The left panel shows the DR13 corrections, which have a large scatter
and average value less than one due to biased
fluxcorr.  The right panel shows the new algorithm in DR14, which reduces scatter and
gives the corrections a mean close to unity.

Due to the highly configurable nature of \texttt{redmonster}, several
input values must be determined for each unique data set to optimize performance.
In this section, we demonstrate the optimization of the most influential, interesting, and challenging of those:
$\Delta \chi_\mathrm{threshold}^2$ and the degree of the polynomial, $n_{\mathrm{poly}}$
to be fit in linear combination
with the templates.  A third important parameter to be determined is $\Delta V$,
the half-width of the window
around a $\chi^2$-minimum in which secondary minima are ignored. This was chosen for eBOSS
as 1000 km~s$^{-1}$ based on clustering science requirements, and thus will not be discussed here.

In order to explore the effects of these parameters and
quantify performance of the software on galaxy spectra, we
analyze the eBOSS LRG target sample, containing 99,449 unique observations of LRG targets.  We use
reductions produced by the tagged version \texttt{v5\_10\_0} of the \texttt{idlspec2d} pipeline
that will be released in DR14.  Determination of 
the configurable $n_{\mathrm{poly}}$ is described in \S\ref{sec:npoly}, and that of 
$\Delta \chi_\mathrm{threshold}^2$ in \S\ref{sec:chi2thresh}.

\subsection{Determination of $n_{\mathrm{poly}}$} \label{sec:npoly}

The number of additive polynomial terms fit in linear combination
with the template spectrum can have a large impact on the quality of the fits and, thus, the 
failure rate and redshift errors.  Some polynomial terms are necessary to absorb broad-band
spectrophotometric calibration errors and astrophysical signal not reflected in the templates. Allowing
too much freedom to the polynomial terms (i.e., too high an order) will allow them to
fit real astrophysical features.  Shifting the quality of the fit to the non-physical components of
the model reduces the ability of the physical templates to provide statistically distinguishable
fits to the data, thus increasing failure rates. In order to understand the effects of $n_{\mathrm{poly}}$,
we processed the LRG sample with \texttt{redmonster}
four separate times, using each of a constant, linear, quadratic, and cubic polynomial, producing
a unique output file for each.

\begin{figure}
\plotone{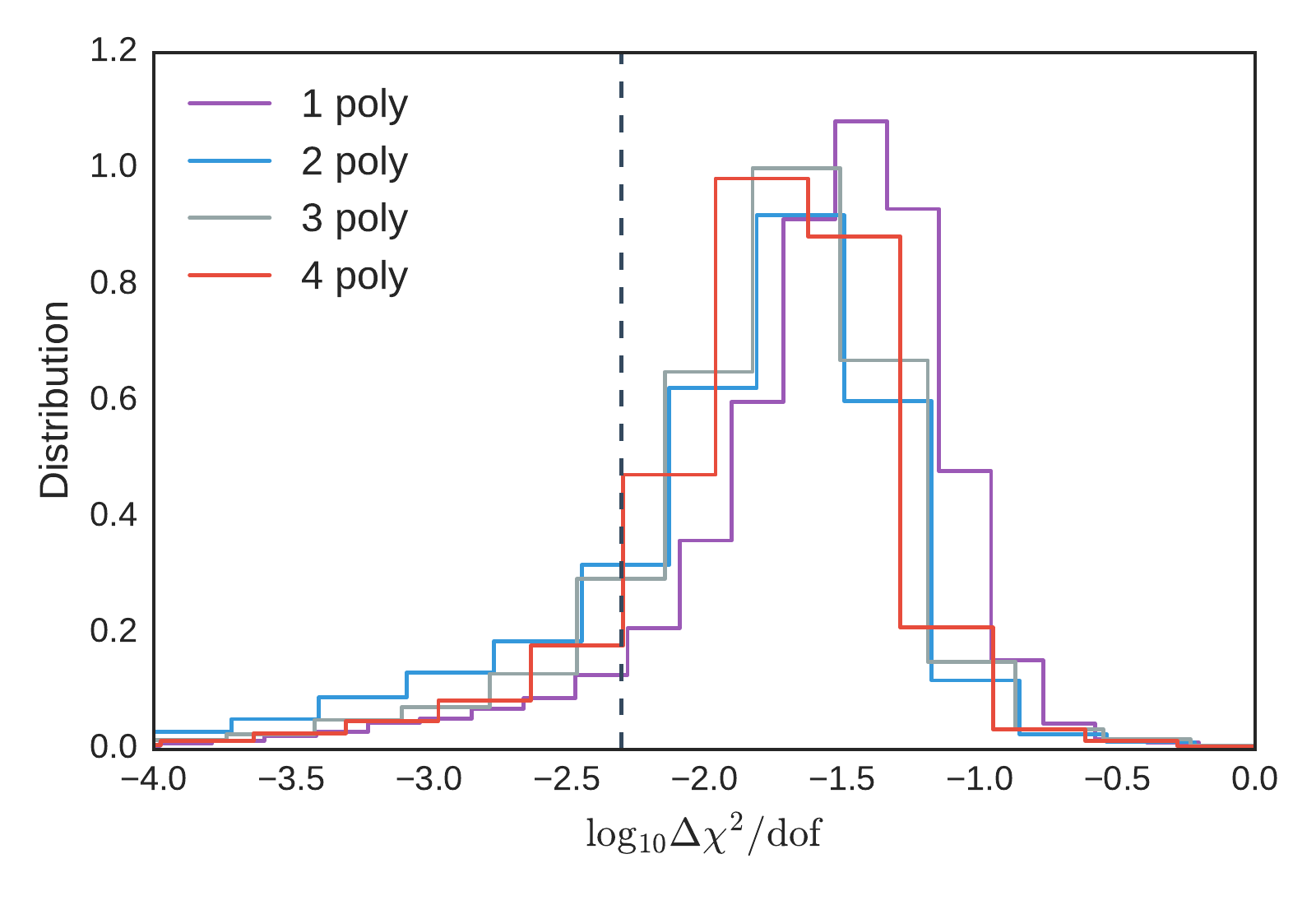}
\caption{Distribution of $\Delta \chi^2$ per degree of freedom of eBOSS LRGs for several
orders of polynomial terms.  The dashed vertical line represents a
$\Delta \chi^2_{\mathrm{threshold}}$ cutoff of 0.005.\\
\label{fig:drchi2_poly_histos}}
\end{figure}

First, we computed the failure rate for each degree of polynomial.
In all cases, the \texttt{ZWARNING} flag corresponding to \texttt{SMALL\_DELTA\_CHI2}
dominates the redshift failure modes.
The failure rates
for a constant, linear, quadratic, and cubic polynomial were 9.5\%, 20.9\%, 14.2\%, and
12.9\%, respectively.
Additionally, we computed the distribution of $\Delta \chi^2$ per degree
of freedom for each run, shown in Figure~\ref{fig:drchi2_poly_histos}.  The use of a constant
polynomial term stands out as having both the lowest failure rate ($\sim 9.5\%$), and a
systematic shift in the $\Delta \chi^2 / \mathrm{dof}$ distribution toward higher values.

In order to understand how the order of the polynomial affects our best-fit models and their
distinguishing power, we focus on the two candidates with the lowest failure rates -
constant and cubic.

\begin{figure*}
\includegraphics[width=\textwidth]{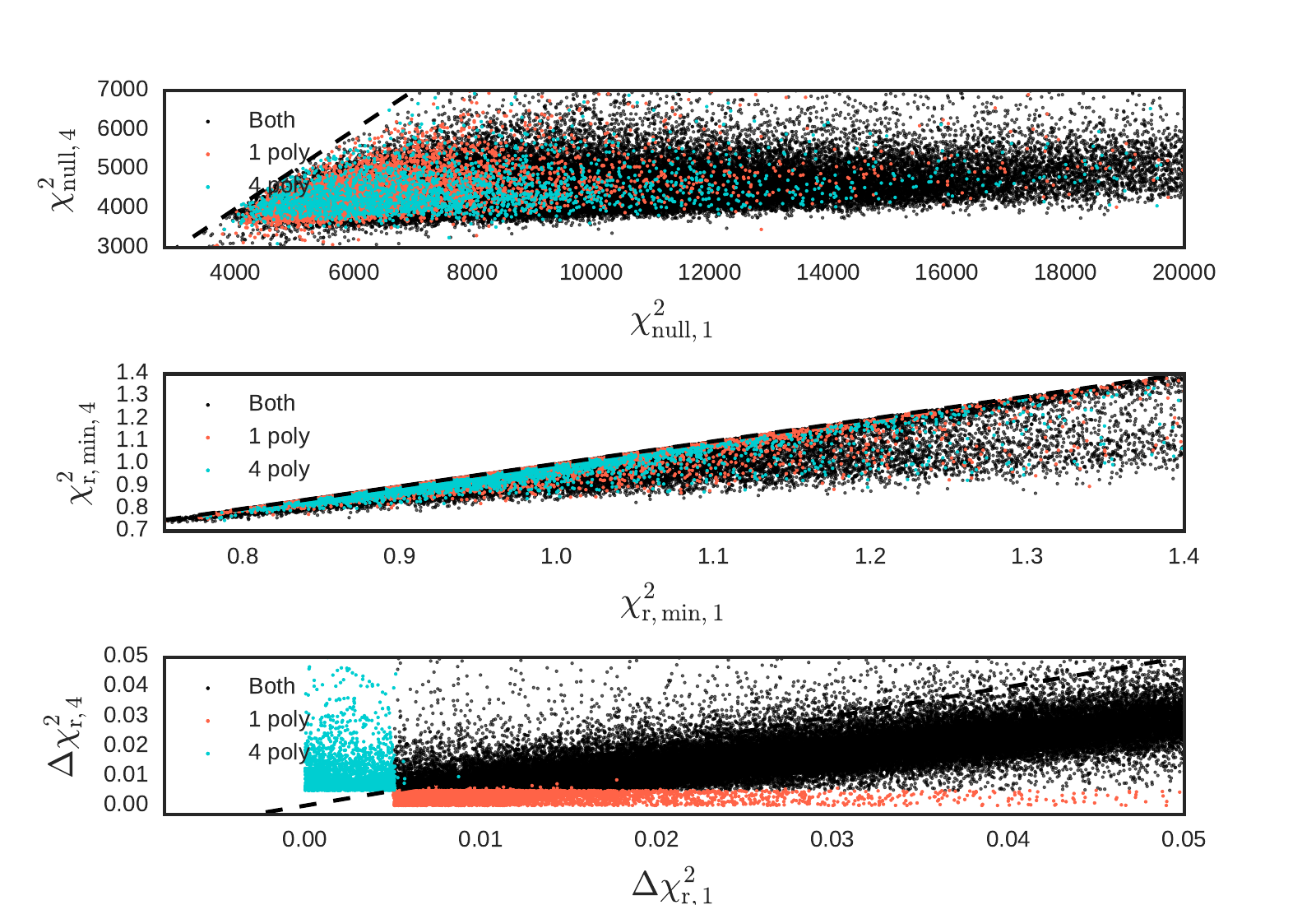}
\caption{Comparisons between several model-fit statistics on an object-by-object basis for
a constant polynomial ($\chi_{~,1}^2$) and cubic polynomial model ($\chi_{~,4}^2$).
The salmon and teal points correspond to
objects where only the constant-polynomial and cubic-polynomial model returned a successful
measurement, respectively.  The dashed line shows the 1-to-1 line in each plot.
Top:  $\chi^2$ of polynomial-only model (i.e., no template component).
Center:  The $\chi^2 / \mathrm{dof}$ of the polynomial + template combination that produces
the best fit to the observed spectrum.
Bottom:  Difference in $\chi^2 / \mathrm{dof}$ between best and second-best model.\\
\label{fig:1poly_4poly_scatters}}
\end{figure*}

Figure~\ref{fig:1poly_4poly_scatters} shows object-matched comparisons of several
statistics of our fits for the constant- and cubic-order polynomials.
We have highlighted cases where one of the models returned
a successful measurement (\texttt{ZWARNING} = 0) while the other failed.
The top panel shows the value $\chi^2_\mathrm{null}$
of each spectrum for both model types.  All points lie to the right of the dashed
line, meaning the constant-polynomial model absorbs less information from the spectrum than
does the cubic-polynomial in every case.  Further, the constant-polynomial has an RMS of $\sim$37,000,
roughly 6.5 times larger than the RMS of the cubic-polynomial.
The narrow range of values around $\chi_\mathrm{null,4}^2=4000-6000$ suggests the cubic-polynomial is consistently 
absorbing most of the broadband information, while the constant-polynomial often cannot.  Meanwhile, the central plot, showing the minimum $\chi^2_{\mathrm{red}}$ for both models, shows a similar range 
of values for each.  Thus, in the cases where the constant-polynomial component is unable to fit
broadband information, the templates effectively take a significant fraction of the signal absorbed
by the cubic-polynomial and transfer it to the physical template.  The bottom panel of
Figure~\ref{fig:1poly_4poly_scatters} illustrates this further.  The data show a linear
relationship between the two values of $\Delta\chi^2 / \mathrm{dof}$ with a slope of $<1$.
On average, $\Delta \chi^2_\mathrm{red}$ is larger for the constant-polynomial model than for the cubic-polynomial model, consistent with the trend in Figure~\ref{fig:drchi2_poly_histos}.

\begin{figure}
\plotone{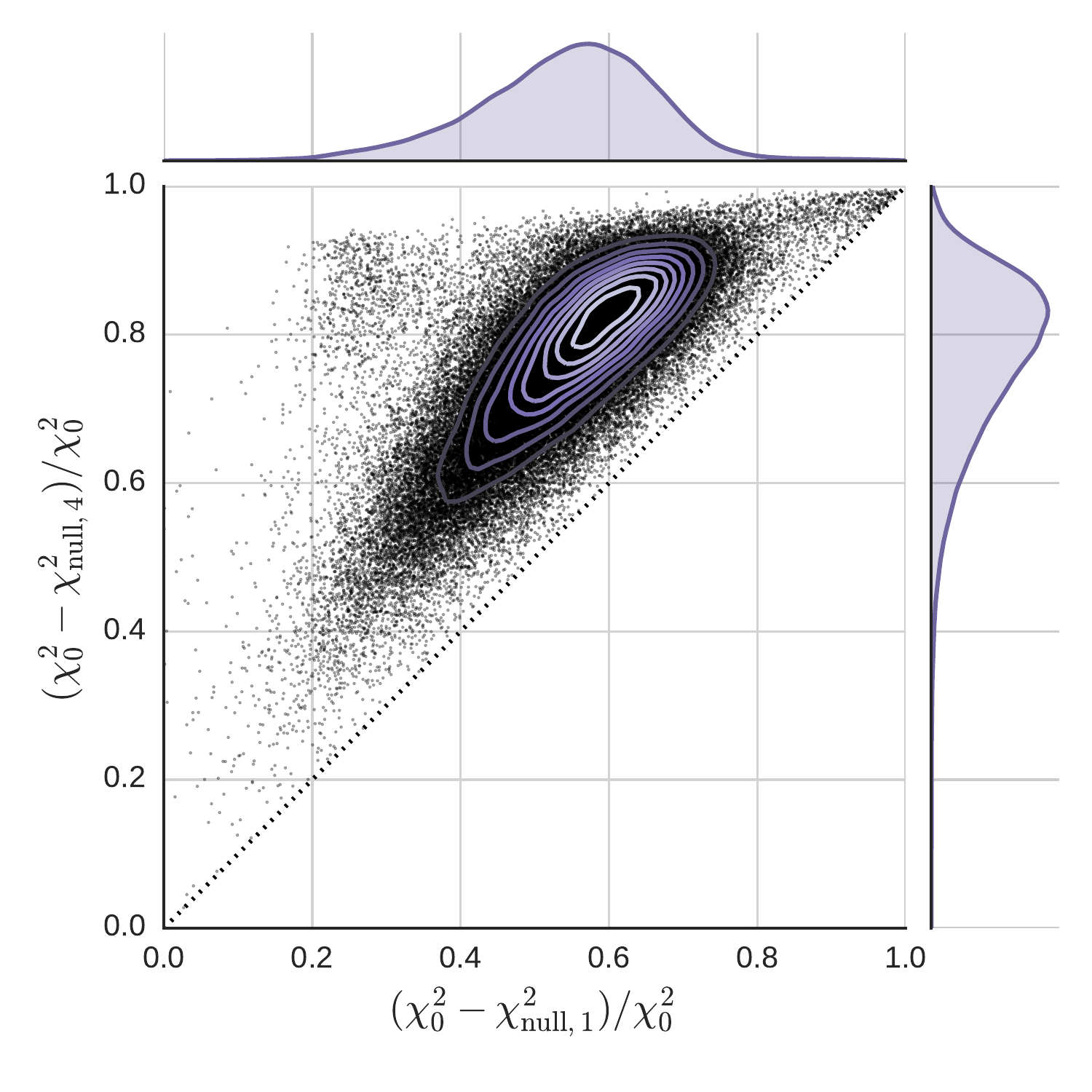}
\caption{Fraction of each object's total (S/N)$^2$ absorbed by the polynomial-only model
for a constant and cubic polynomial, with bivariate kernel density estimate overlaid.
The dotted line shows the unity relationship.
Marginal plots show univariate kernel density estimates for each axis.\\
\label{fig:jointplot3}}
\end{figure}

Figure~\ref{fig:jointplot3} shows
$(\chi^2_0 - \chi^2_{\mathrm{null}})/\chi^2_0$ of both polynomial choices for each object,
where $\chi^2_0$ is the $\chi^2$ of a zero model (i.e, the (S/N)$^2$ of the data); thus,
this quantity may be interpreted as the
fraction of the total (S/N)$^2$ of the data absorbed by the polynomial.
Note that all data points fall above the unity relationship, meaning the cubic polynomial
always absorbs more of the information in the spectrum than does the constant.
We have overlaid contours from a bivariate kernel
density estimate, which has a maximum at (0.573, 0.831).
The marginal plots show the univariate kernel density estimates for the constant polynomial
on the horizontal axis and cubic polynomial on the vertical axis. The median values of the
constant and cubic are 0.553 and 0.787, respectively, suggesting that, on average, the cubic
polynomial and spectral template absorbs $\sim 23\%$ more, in absolute terms,
of the spectrum's signal than does the constant polynomial and spectral template.

\begin{figure*}
\plotone{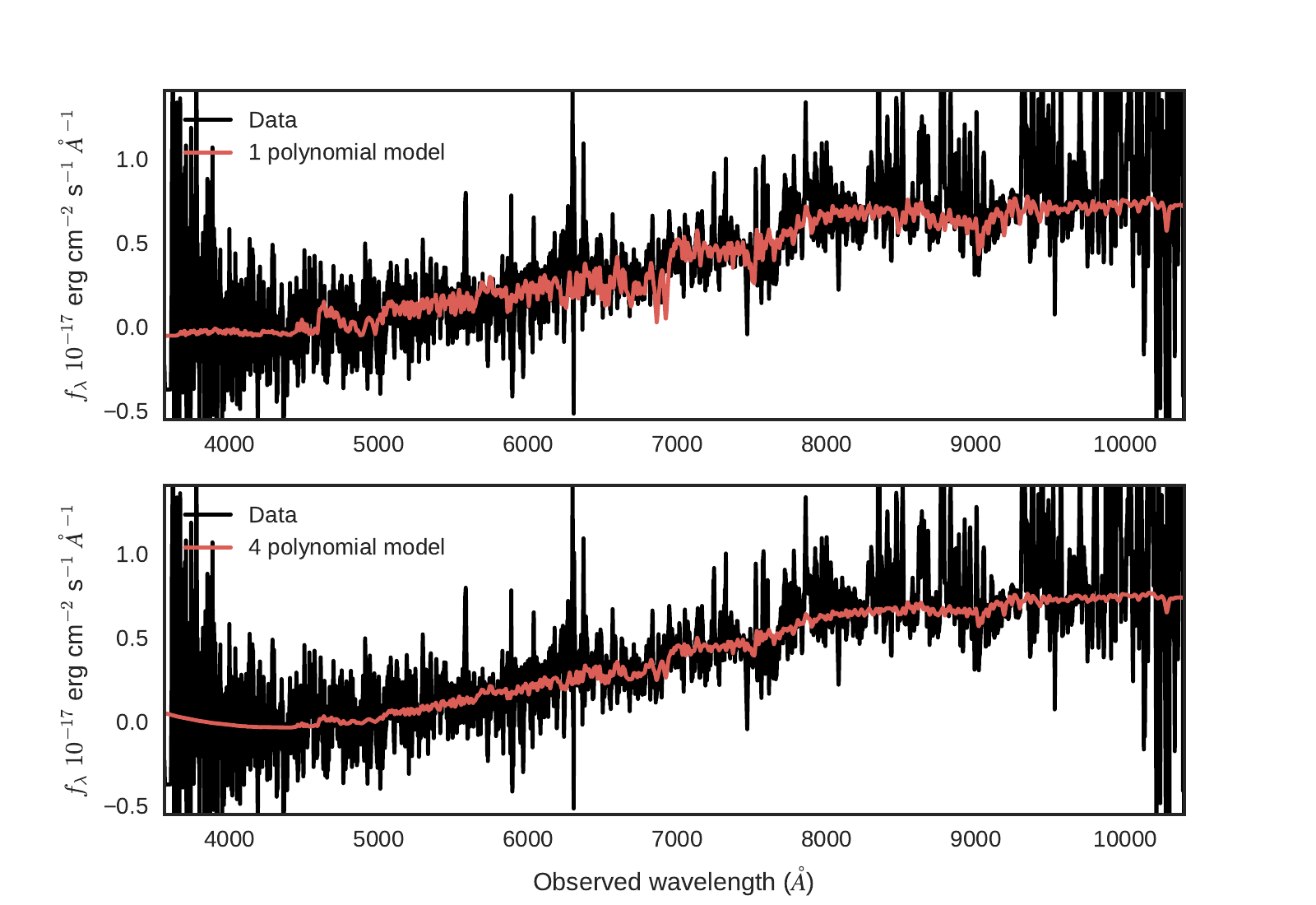}
\caption{Example of an object (\texttt{PLATE} 7572 \texttt{MJD} 56944 \texttt{FIBERID} 515)
in which a constant-polynomial model returned a successful
measurement while the cubic-polynomial measurement failed.  The data are shown
in black, and the best-fit model is shown in red.
\\
\label{fig:yes1no4}}
\end{figure*}

As a final means of exploring the effects of changing the order of the polynomial, we visually
inspected spectra which returned a successful measurement in one run but a failure in the other.
The visual inspections help us to develop a qualitative understanding of the types of
spectra or spectral features that are
handled successfully in one case and are problematic for the other.  Figure~\ref{fig:yes1no4} shows
a $z = 0.743$ galaxy from the LRG sample for which the constant-polynomial model returned
a successful redshift measurement and the cubic-polynomial model returned a failure.  This
spectrum was chosen as being representative of the typical behavior of models with
constant-success and cubic-failure.  There is a clear unphysical upturn in the noisy blue end of the spectrum
due to calibration errors, which happens occasionally among low S/N eBOSS spectra.  The higher
flexibility of the cubic-polynomial fit allows the model to chase this upturn, which shifts
power into the non-physical component of the model and away from the physical template.
Because the long wavelength corrections are coupled to the short wavelength corrections through
the cubic polynomial, 
this results in the suppression of the equivalent width of narrow-band features such as Ca~H\&K
at 3968.5 and 3933.7~\AA, respectively, H$\beta$ at 4863~\AA,
Mg~I at 5175~\AA, and Na~I at 5894~\AA.
In low signal-to-noise spectra, such as those of eBOSS, it becomes difficult for the software to
distinguish between the model fitting a real narrow-band feature and fitting noise.
In this case, the cubic-polynomial
model does, in fact, have the correct redshift, but due to reduced template amplitude relative to
the constant-polynomial model, lacks the statistical confidence to declare a confident measurement.

\begin{figure*}
\plotone{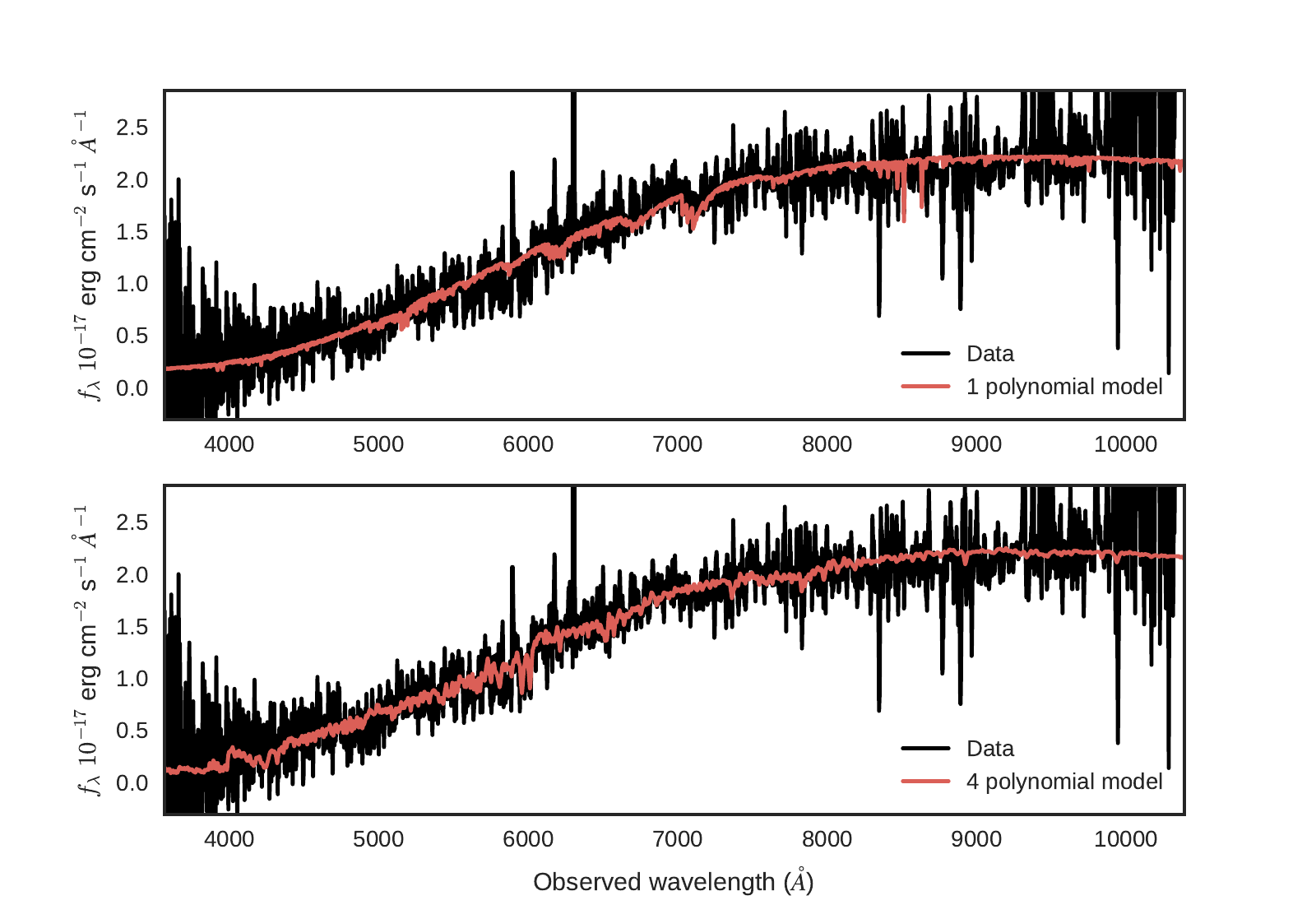}
\caption{Example of an object (\texttt{PLATE} 7575 \texttt{MJD} 56947 \texttt{FIBERID} 434)
in which a cubic-polynomial model returned a successful
measurement while the constant-polynomial measurement failed.  The data are shown
in black, and the best-fit model is shown in red.
\\
\label{fig:no1yes4}}
\end{figure*}

On the other hand, for spectra with the largest broadband deviations from the templates
or deviations that extend beyond the blue end (due to
flux calibration errors, object superpositions, etc.), the constant-polynomial model lacks the
flexibility to fit these features.  In these cases, the $\chi^2$ is driven by poor fitting of the broadband
features, and the relative contribution from the astrophysical features is small.  This hinders the
software's ability to distinguish between models with different physical parameters.  The
cubic-polynomial, on the other hand, has the flexibility to absorb these strong broadband features,
allowing the astrophysical features to dominate the $\chi^2$, and the software is able to return a 
successful measurement.  An example of this type of spectrum is shown in Figure~\ref{fig:no1yes4},
a $z = 0.513$ galaxy in the LRG sample
in which broadband features extend across the entire range of the spectrograph.
The constant-polynomial is unable to capture this, forcing the fitter to choose a stellar template, despite
clear absorption features being visible from Ca~H\&K around 6000~\AA,
the G-band around 6500~\AA, H$\beta$ around 7350~\AA, and
a less well-defined Mg~I line around 7800~\AA.
The cubic-polynomial was able to absorb the broadband
features, meaning the $\chi^2$ is dominated by the template's fit to astrophysical signal and
a successful measurement was returned.

At this point, it is clear that a constant-polynomial term produces lower failure rates, and is
empirically the best choice for our data set.  Further, a qualitative understanding
of the failure modes of the two model types leads us to believe that, in the presence of better
flux calibration, the failure rates of a constant-polynomial are likely to decrease further.
Thus, we use a constant-polynomial model
to classify the eBOSS LRG sample and in all subsequent analyses.

\subsection{Determination of $\Delta \chi_\mathrm{threshold}^2$} \label{sec:chi2thresh}

Decreasing the value of $\Delta \chi_\mathrm{threshold}^2$
relaxes the requirement for a declaration of statistical confidence, resulting
in higher redshift completeness rates. Doing so comes at the expense of
higher rates of catastrophic failures,
as incorrect measurements that would have been flagged at a higher threshold are allowed
through with no \texttt{ZWARNING} flag.  Similarly, increasing the value of
$\Delta \chi_\mathrm{threshold}^2$ decreases catastrophic failure rates as it
restricts statistical confidence to only the best of fits, but does so at the expense of
completeness rates, reducing usable sample size and statistical power towards cosmology constraints. 
Thus, determining a value of $\Delta \chi_\mathrm{threshold}^2$ means striking
an optimal balance between
completeness and purity while ensuring science requirements are met.  Here, we remind the reader
that $\Delta \chi_\mathrm{threshold}^2$ is always scaled to the degrees of freedom
to account for possibly varying degrees of freedom between fits, where the number of
degrees of freedom is defined as the number of unmasked pixels in the spectrum less the number
of template components (i.e., $n_\mathrm{pix} - (n_\mathrm{poly}+1)$).

\begin{figure}
\plotone{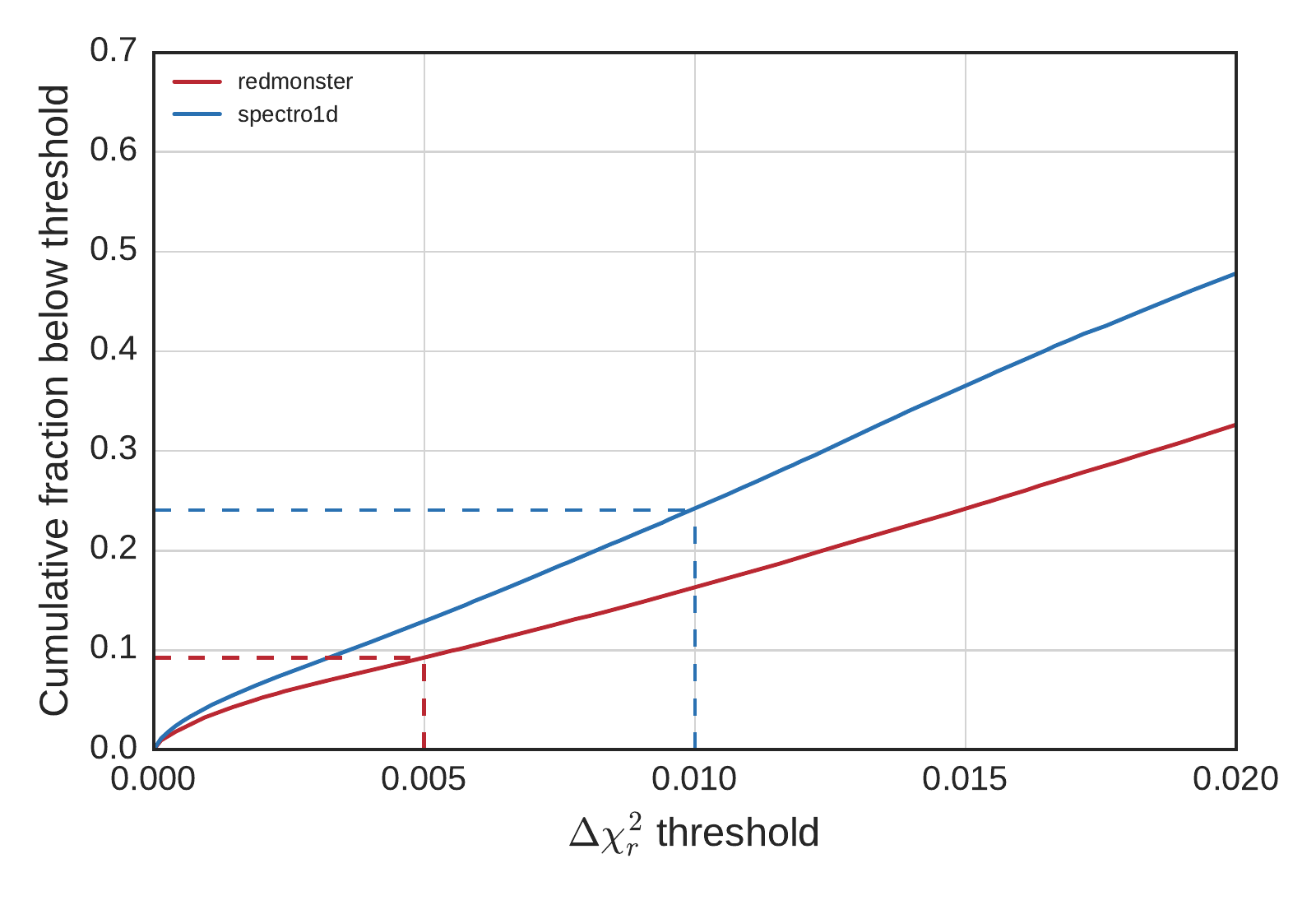}
\caption{Redshift failure (\texttt{ZWARNING} $> 0$) rate of \texttt{redmonster} and \texttt{spectro1d}
pipelines as a function of $\Delta\chi_\mathrm{r,threshold}^2$to trigger the \texttt{SMALL\_DELTA\_CHI2}
bit.  Dashed red and blue lines serve to visually illustrate the failure rates of \texttt{redmonster}
and \texttt{spectro1d} at $\Delta\chi_\mathrm{r,threshold}^2$ values of 0.005 and 0.01, respectively.
\\
\label{fig:drchi2_vs_failure}}
\end{figure}

We first investigate
the failure rate as a function of $\Delta \chi_\mathrm{threshold}^2$, as shown in
Figure~\ref{fig:drchi2_vs_failure}.  At
$\Delta \chi_\mathrm{threshold}^2 = 0.01$, the value used by \texttt{spectro1d}, \texttt{redmonster}
reduces the failure rate from 24.3\% to 16.3\%. 
While significant, the failure rate is still more than a factor of 1.5 times
the desired $\le10$\% failure rate for eBOSS science.  To ensure sub-10\% failure rates,
the \texttt{ZWARNING} flag for \texttt{SMALL\_DELTA\_CHI2} would need to be set
at or below $\Delta \chi_\mathrm{threshold}^2 = 0.005$.

While quantifying completeness is straightforward, doing so for catastrophic failure rates is not.
The eBOSS science requirement for catastrophic failures is $<1\%$.
By definition, catastrophic failures pass through the software unnoticed, 
making identification difficult. We consider two tests to characterize the rate of catastrophic failures.

\begin{figure*}[h!]
\plottwo{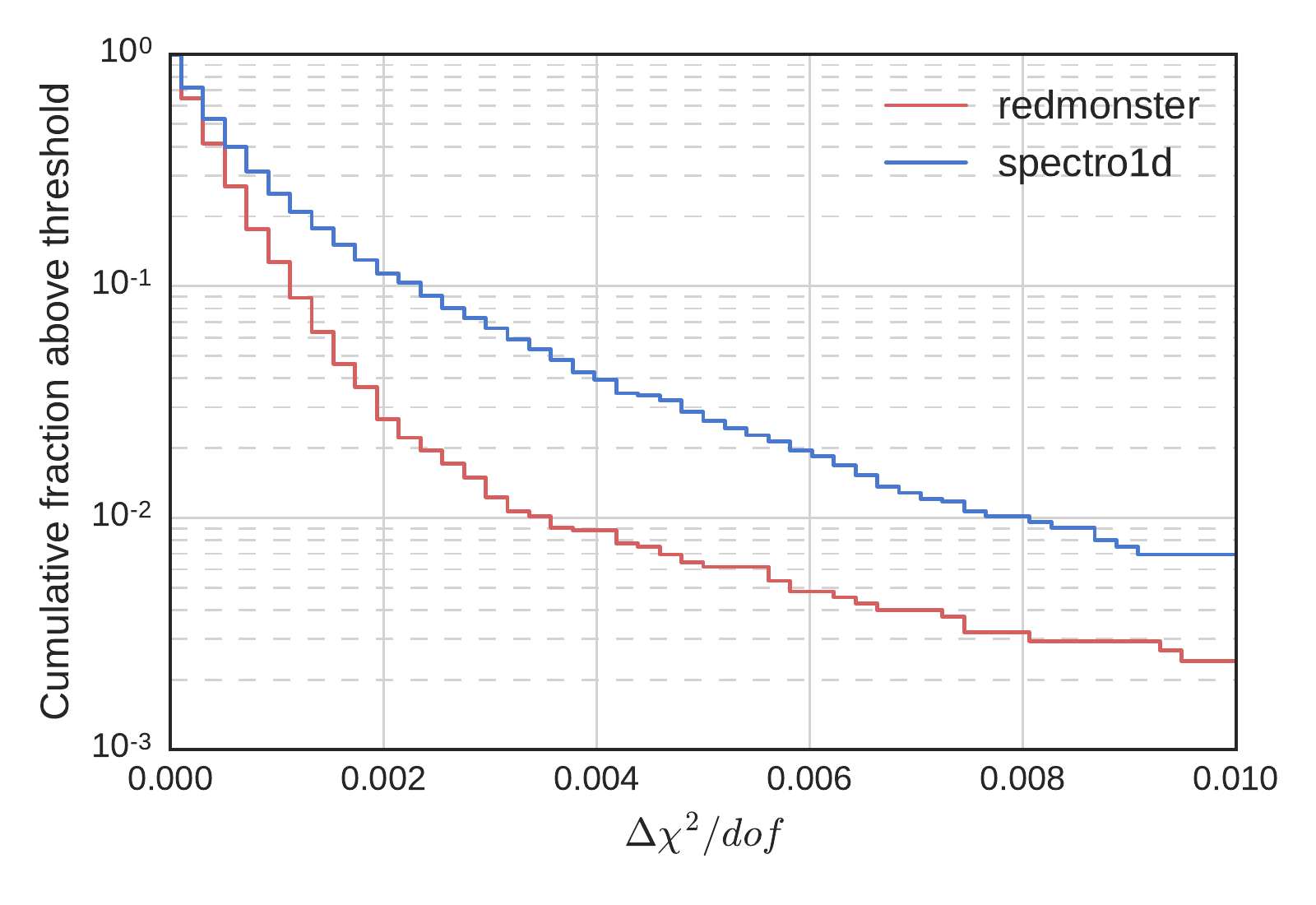}{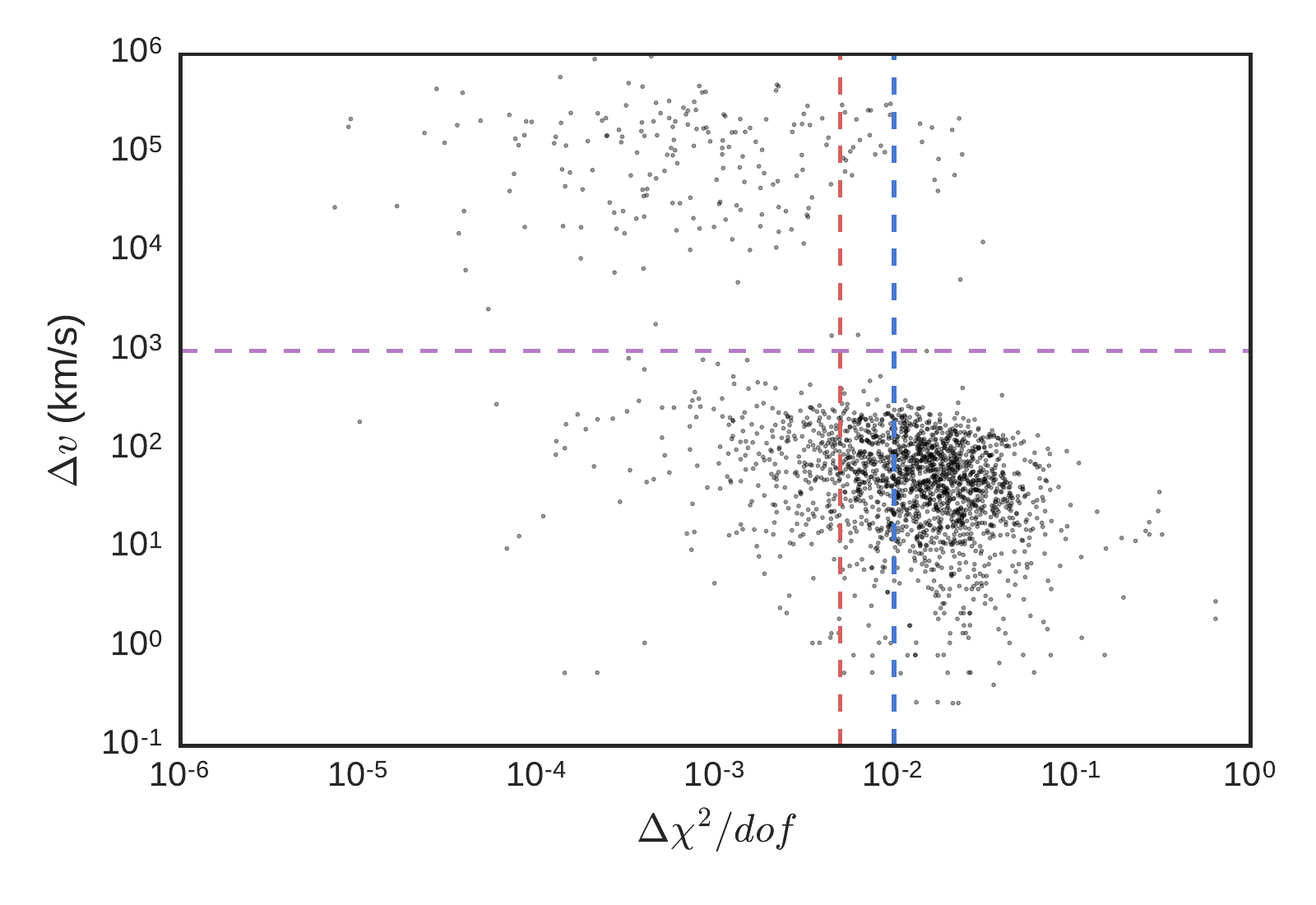}
\caption{Left:  Cumulative fraction of eBOSS sky fibers with confident redshift measurement and
classification as a function of $\Delta\chi_\mathrm{threshold}^2$.
Right:  Scatterplot showing redshift difference (in km s$^{-1}$) between independent observations
of the same LRG target.  The red and blue vertical lines represent $\Delta \chi^2 / \mathrm{dof}$
thresholds of 0.005 and 0.01, respectively.  The horizontal dashed line shows the limit of
velocity difference to be considered a catastrophic failure.\\
\label{fig:sky_failure_vs_drchi2}}
\end{figure*}

We first asses the completeness of eBOSS sky fibers as a function of
$\Delta \chi_\mathrm{threshold}^2$
as a proxy for catastrophic failures.
Sky fibers are those fibers that are not placed on any target, but rather are intended to measure
the sky emission, unpolluted by astronomical objects, to aid in sky-subtraction.  Roughly
$\sim 10\%$ of the fibers on each eBOSS plate are placed as sky fibers.
While a small fraction of these will inevitably be placed over real
objects, the vast majority should contain no object at all.  These should trigger the softwares'
\texttt{SMALL\_DELTA\_CHI2} flag since there is no astronomical object; a confident redshift
is impossible.  While the rate of false positives within this sample is not a direct measurement
of the catastrophic failure rate, it does provide a testing ground for the software's behavior in the limit
of low signal-to-noise.  As signal-to-noise is the best predictor of redshift measurement success,
this sample is informative of the true rates of catastrophic failures.
The left panel of Figure~\ref{fig:sky_failure_vs_drchi2} shows the cumulative
fraction of eBOSS sky fibers above a given $\Delta \chi_\mathrm{threshold}^2$ for \texttt{redmonster}
and \texttt{spectro1d}.  At a given threshold,
\texttt{redmonster} returns a factor of seven fewer confident measurements.
The increased
rigidity of modeling with a single, physically-motivated template over a set of PCA basis vectors
reduces the ability of \texttt{redmonster} to fit sky residuals and noise, greatly reducing its
rate of false positives.  The eBOSS science requirement for catastrophic failures is $<1\%$,
which we see in \texttt{redmonster} at a $\Delta \chi_\mathrm{threshold}^2$ value of $\sim~0.004$.
In \texttt{spectro1d}, meanwhile, it does not reach sub-1\% values until $\sim0.008$.

We then assess the redshift differences between different spectra of the same object,
as in \citet{daw2016}.  First, we identified 2128 targets that were tiled on more than one plate
and, thus, have multiple independent observations.  We then compared redshift differences
as a function of $\Delta \chi^2 / \mathrm{dof}$, as shown in
the right panel of Figure~\ref{fig:sky_failure_vs_drchi2}.
Assuming that, in the cases of discrepant redshifts, one redshift is correct, the rate of catastrophic
failures can be estimated by counting objects with $\delta z>1000$ km s$^{-1}$ and
$\Delta \chi^2 / \mathrm{dof}$ above the threshold.
For this sample, there are 12 catastrophic failures out of 2520 confident measurements
for $\Delta\chi_\mathrm{threshold}^2 = 0.01$ and
32 catastrophic failures out of 3270 confident measurements for $\Delta\chi_\mathrm{threshold}^2 = 0.005$,
corresponding to catastrophic failure rates of $0.48\pm 0.14$\% and
$0.98\pm 0.22$\%, respectively.  A similar analysis for the \texttt{spectro1d} reductions
yields failure rates of $0.35\pm 0.11$\% and $0.92\pm 0.16$\%.


In order to maximize completeness while maintaining an acceptable catastrophic failure
rate, we set $\Delta\chi_\mathrm{threshold}^2=0.005$ for all subsequent analyses.
We also use $\Delta\chi_\mathrm{threshold}^2=0.01$ for \texttt{spectro1d} when making
comparisons, to ensure that the catastrophic failure rate requirement is being met in both sets of reductions.
Amongst the full set of eBOSS LRG target spectra, we find an automated completeness
(\texttt{ZWARNING} == 0) rate of 90.5\%, with a catastrophic failure rate of 0.98\%.
Meanwhile, \texttt{spectro1d} produces a completeness of 75.6\% with a
catastrophic failure rate of 0.32\%.
Thus, \texttt{redmonster} satisfies the requirements of completeness and purity, while \texttt{spectro1d}
does not.
This improvement is illustrated by the dashed lines in Figure~\ref{fig:drchi2_vs_failure}.

\section{Classification of LRG spectra from \MakeLowercase{e}BOSS} \label{sec:classification}

We made use of 99,449 eBOSS LRG targets reduced with tagged version \texttt{v5\_10\_0}
of \texttt{idlspec2d} to demonstrate the performance of \redmonster.
Galaxy redshift success dependence is described in \S\ref{sec:dependence}, 
the effect on the final LRG sample redshift distribution is described in
\S\ref{sec:distribution}, and galaxy redshift precision and accuracy
in \S\ref{sec:precision}.  A description of composite spectra and the distribution
of physical galaxy parameters in the sample is given in \S\ref{sec:composites}.

\subsection{Galaxy redshift success dependence} \label{sec:dependence}

\begin{figure*}
\plottwo{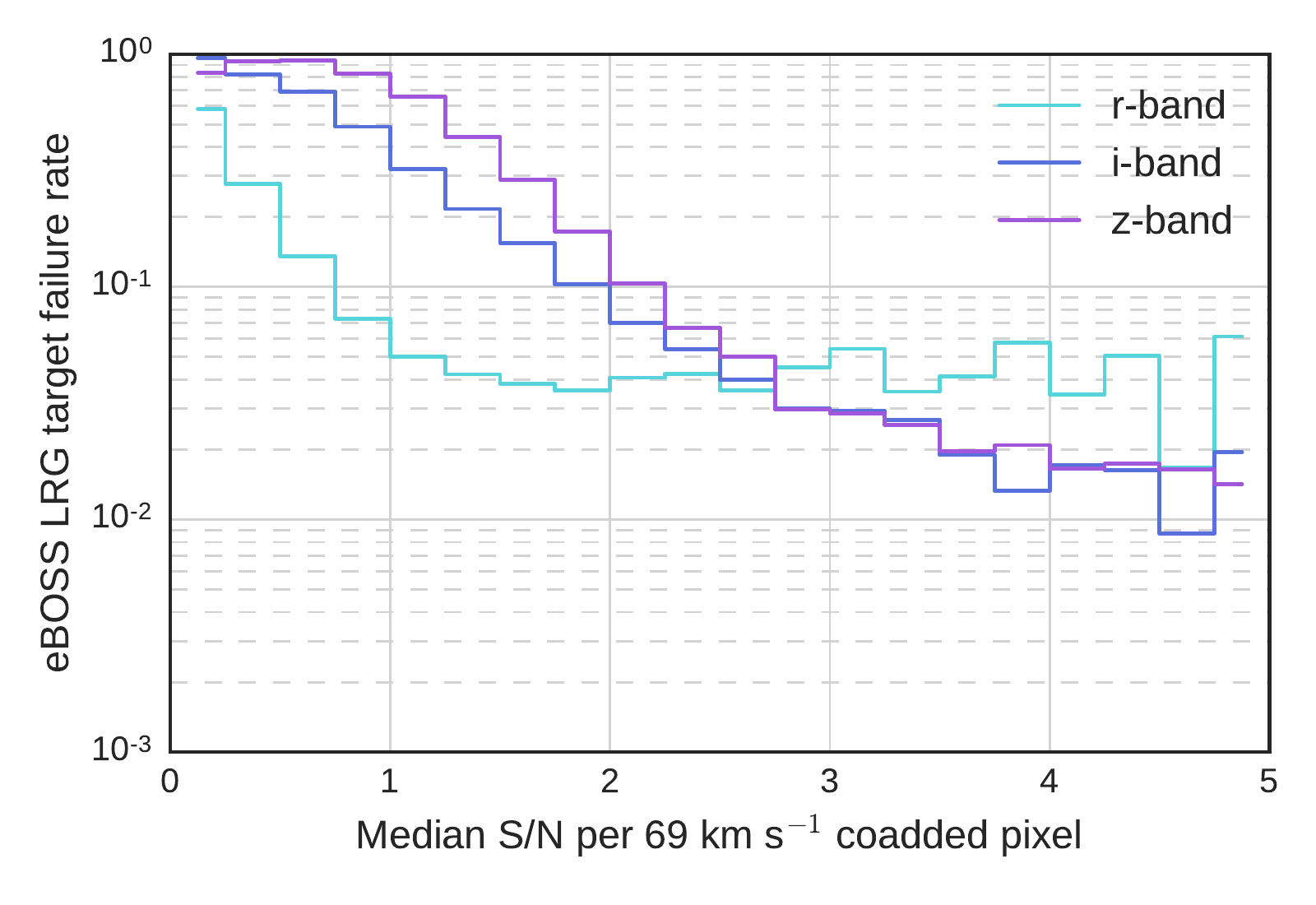}{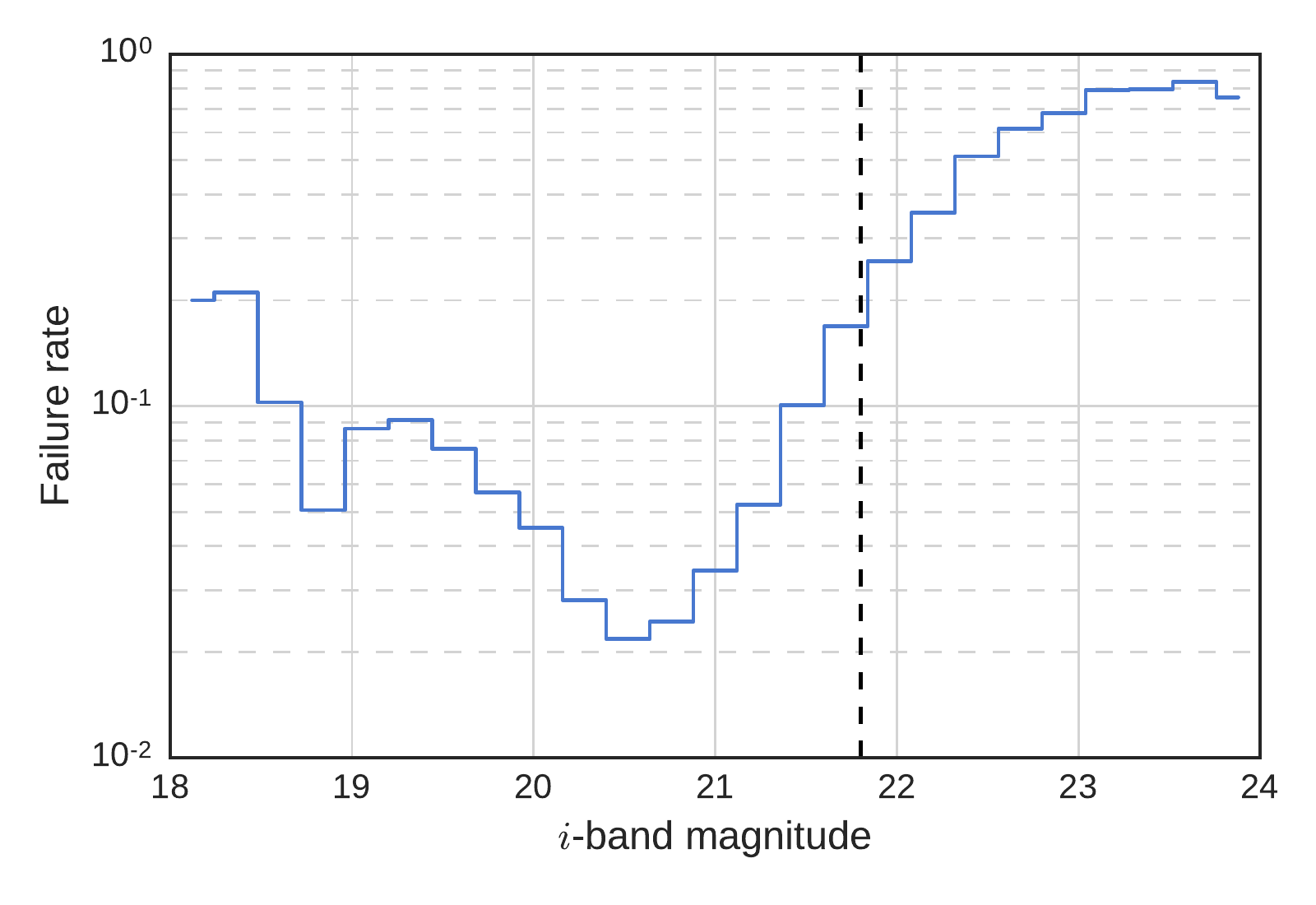}
\caption{Left:  Redshift failure (\texttt{ZWARNING} $> 0$) rate of the eBOSS LRG sample as a
function of median S/N in SDSS $r$-, $i$-, and $z$-bands.  Right:  Redshift failure
(\texttt{ZWARNING} $> 0$) rate of the eBOSS LRG sample as a function of SDSS
$i$-band magnitude.  The dashed vertical line indicates the faint-magnitude limit of $i=21.8$.\\
\label{fig:failure_vs_sn}}
\end{figure*}

As in all redshift surveys, spectroscopic S/N is the primary determinant of redshift success.
In the eBOSS LRG sample, 95.4\% of spectra with \texttt{ZWARNING} $>$ 0 are due
solely to a \texttt{SMALL\_DELTA\_CHI2} failure.
Figure~\ref{fig:failure_vs_sn} shows the dependence of the LRG galaxy redshift failure rate as a function of
the median spectroscopic signal-to-noise ratio over the SDSS r, i, and z bandpass ranges. These represent
the most relevant regions of the spectrum for measuring redshifts of passive galaxies over the redshift range
of interest for the large-scale structure science in eBOSS. Failure is defined in the sense of
\texttt{ZWARNING} $> 0$, so that targets confidently identified as objects other than galaxies are counted
as a success for the pipeline.  We see a decrease in the failure rate as a function of r-band S/N up to S/N 
$\sim 1.8$, where it becomes asymptotic to $\sim 3\%$.
The $i$- and $z$-bands behave in a more expected manner, with failure rate
decreasing until S/N $\sim 4$, where it reaches an asymptotic minimum of $\sim 2\%$.
The $i$- and $z$-band S/N is more predictive of redshift success rate due to the
4000~\AA\ break and the small
number of strong narrow absorption features (e.g., Ca H\&K, Na I, etc.)
being located in those bands over the targeted
redshift range.

Galaxy magnitude correlates strongly with spectroscopic S/N and hence with redshift success; this is the
motivation for the $i$-band magnitude limit of $< 21.8$ in the target selection algorithm.
To assess the dependence of redshift completeness
on target selection, the right panel of Figure~\ref{fig:failure_vs_sn} shows the LRG sample's
redshift failure rate as a function of
$i_{\mathrm{fiber}}$, defined as the $i$-band magnitude in a 2\arcsec~diameter eBOSS fiber.
At the formal eBOSS LRG magnitude cutoff, the marginal failure
rate is $\sim~11\%$.

\begin{figure*}
\plottwo{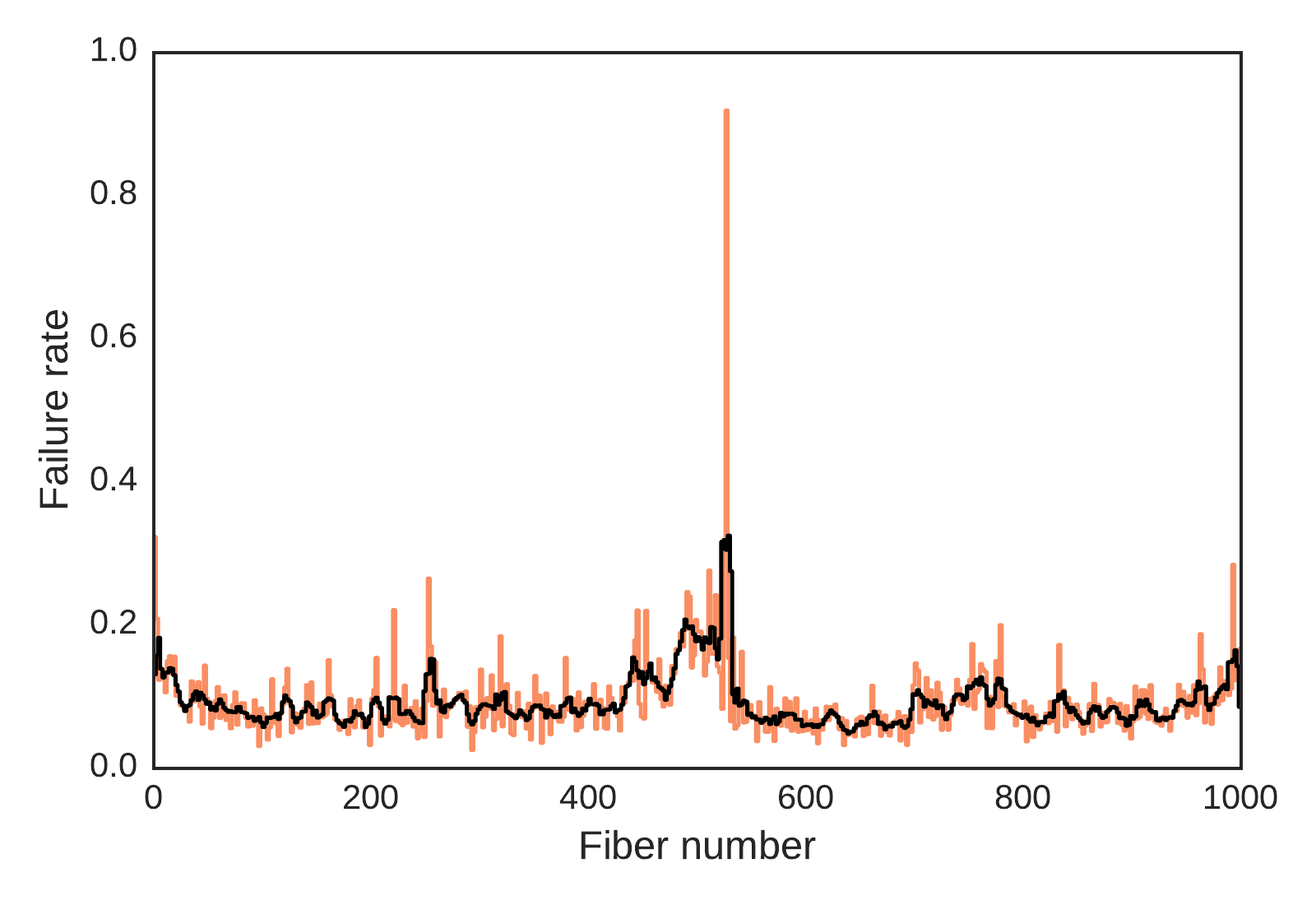}{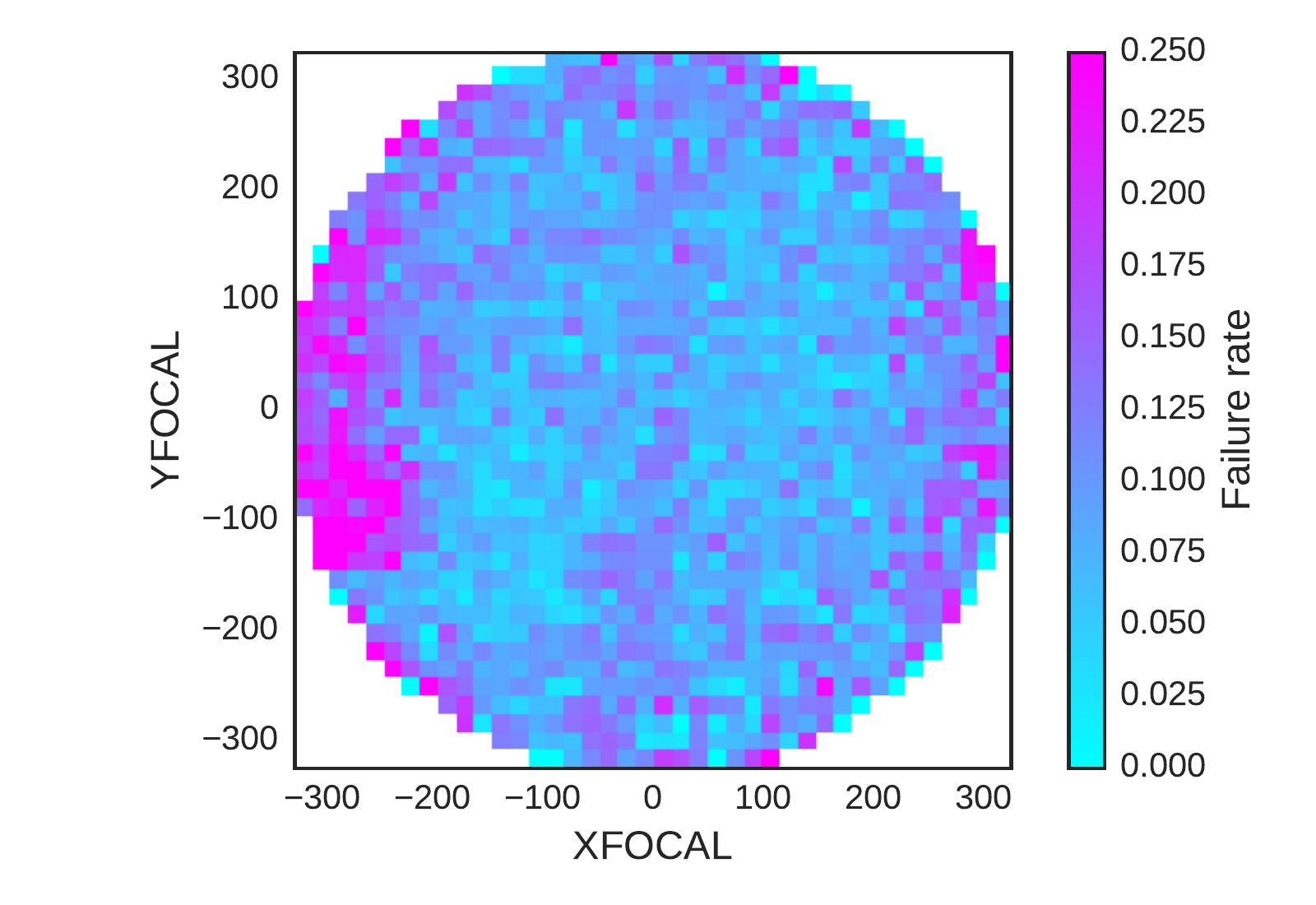}
\caption{Left:  eBOSS LRG sample failure rate as a function of fiber number.  The orange plot
shows individual fibers, while the black has been smoothed by 5 pixels to highlight large-scale
structure.
Right:  Failure rate of eBOSS LRG sample as a function of location of fiber head on the
physical plug plate.\\
\label{fig:failure_vs_fiberid}}
\end{figure*}

Additionally, redshift success has a weak dependence on fiber identification
number along the spectrograph slit heads.
The left panel in Figure~\ref{fig:failure_vs_fiberid} shows this effect for eBOSS LRGs.
Upturns near fibers 1, 500 and 1000
are due to imperfections in the camera optics near the edge of the spectrograph focal plane.
Narrow peaks, such as those around fibers 525-530, are due to bad CCD columns.
Fiber numbers below 500 show a higher average failure rate (9.4\%) than those above 500 (9.0\%)
due to lower
end-to-end throughput of spectrograph 1 relative to spectrograph 2 \citep{sme2013}.

Finally, we investigated the dependence of failure rate on the location of the fiber on the plug plate.
The right panel of
Figure~\ref{fig:failure_vs_fiberid} shows this relationship.  The fibers within the central region
covering 50\% of the total area show failure rates of $7.8\%$. However, near the edges and, particularly,
the left and right sides of the plate, failure rates spike to values of 25\% or greater.  Plates are generally
plugged in a counter-clockwise direction, beginning in the first quadrant, meaning the right edge of the plate
contains fibers near 1 and 1000, while the left edge contains fibers near 500; thus, the increased
failure rates at the extreme values of XFOCAL are primarily due to the imperfect spectrograph
optics described above.


\subsection{Effect on final redshift distribution and cosmological projections} \label{sec:distribution}

In order to quantify the effects of the \texttt{redmonster} spectral classification
on the survey's expected cosmological constraints, we compare the resulting redshift
distribution to the predictions presented in \citet{daw2016}.  Those cosmological projections
were based on the redshift distribution derived from
visually inspected redshifts and classifications
for 1997 LRG targets across 16 plates.
Plates with deeper than average observations were intentionally
chosen to facilitate these visual inspections.
These spectra were 
processed using \texttt{idlspec2d} tagged version \texttt{v5\_8\_0}, and were visually inspected
by ten members of the eBOSS team in August 2015.  Each spectrum was
manually assigned a redshift and spectral classification, as well as a confidence
value $q_\mathrm{conf}$ ranging from 0 (entirely uncertain) to 3 (entirely certain).
A full qualitative description of all $q_\mathrm{conf}$ values is given in \citet{daw2016}.
A subset of plates was inspected
by two people, providing a degree of self-calibration of the results.


\begin{deluxetable*}{lcccccc}
\tabletypesize{\footnotesize}
\tablecolumns{7}
\tablewidth{0pt}
\tablecaption{\label{tab:n_of_z} Redshift distribution for the LRG sample from visual
inspections, \texttt{spectro1d}, and \texttt{redmonster} using tagged versions \texttt{v5\_9\_0}
and \texttt{v5\_10\_0} of the reductions. The surface densities are presented in units of
deg$^{-2}$ assuming that 100\% of the objects in the parent sample are spectroscopically
observed. Entries highlighted in bold font denote the fraction of the sample that
satisfies the high-level requirement for the redshift distribution of the sample.}
\tablehead{
\colhead{} & \colhead{Visual} & \colhead{Visual} &
\colhead{\texttt{spectro1d}} & \colhead{\texttt{redmonster}}
& \colhead{\texttt{spectro1d}} & \colhead{\texttt{redmonster}}\\
\colhead{} & \colhead{($q_\mathrm{conf}>0$)} & \colhead{($q_\mathrm{conf}>1$)} &
\colhead{(DR13)} & \colhead{(DR13)} & \colhead{(DR14)} &
\colhead{(DR14)}
}
\startdata
Poor Spectra & 4.0 & 6.7 & 17.7 & 7.7 & 14.3 & 5.7\\
Stellar & 5.3 & 5.3 & 6.5 & 2.9 & 5.4 & 4.9\\
$0.0<z<0.5$ & 0.6 & 0.6 & 0.6 & 0.5 & 0.6 & 0.5\\
$0.5<z<0.6$ & 6.2 & 5.9 & 5.2 & 6.2 & 3.4 & 6.2\\
$0.6<z<0.7$ & \bf{15.2} & \bf{14.8} & \bf{11.3} & \bf{14.3} & \bf{12.4} & \bf{14.7}\\
$0.7<z<0.8$ & \bf{15.3} & \bf{14.7} & \bf{11.4} & \bf{16.1} & \bf{12.9} & \bf{15.3}\\
$0.8<z<0.9$ & \bf{9.4} & \bf{8.7} & \bf{5.6} & \bf{8.8} & \bf{6.8} & \bf{8.9}\\
$0.9<z<1.0$ & \bf{3.2} & \bf{2.7} & \bf{1.4} & \bf{2.5} & \bf{1.7} & \bf{2.7}\\
$1.0<z<1.1$ & 0.5 & 0.4 & 0.3 & 0.9 & 0.3 & 1.0\\
$1.1<z<1.2$ & 0.1 & 0.1 & 0.1 & 0.2 & 0.1 & 0.2\\
\hline\\
Total Targets & 60 & 60 & 60 & 60 & 60 & 60\\
Total Tracers & \bf{43.1} & \bf{41.0} & \bf{29.7} & \bf{41.7} & \bf{33.9} & \bf{41.7}
\enddata
\end{deluxetable*}

The survey science goals require a minimum of 40 deg$^{-2}$ spectroscopically confirmed LRGs
in the redshift range $0.6<z<1.0$.
To compensate for incomplete fiber assignment, the LRG parent sample
is selected at a surface density of 60 deg$^{-2}$.  Table~\ref{tab:n_of_z} shows the
estimated final tracer density of the LRG sample, binned by redshift.  We show the optimistic
($q_\mathrm{conf}>0$) and conservative ($q_\mathrm{conf}>1$) scenarios
from the visual inspections alongside the \texttt{redmonster} and \texttt{spectro1d}
N$(z)$ results using reduction versions \texttt{v5\_9\_0} and \texttt{v5\_10\_0}.
The surface density of tracers is increased by \texttt{redmonster} relative to \texttt{spectro1d}
by 40.4\% and 23.9\% in DR13 and DR14, respectively.  While the improvements to the
reductions described in \S\ref{sec:parameters} increased the rate of successful redshifts
by \texttt{spectro1d} by 14.1\%, the tracer density for \texttt{redmonster} remained constant at
41.7~deg$^{-2}$. Twenty-six percent of spectra that were previously failures were reclassified
by \texttt{redmonster} as stars
in the improved reductions.
In the optimistic case using extra deep spectra,
the visual inspections report a failure of $\sim6.7$\%, while \texttt{redmonster} finds a failure rate
of 7.4\% on those same spectra.
This suggests \texttt{redmonster} is producing failure rates nearly as low as what any
software can achieve.

We use the final redshift distribution from \texttt{redmonster} to predict changes to the cosmological
projections given in \citet{daw2016}.  Those projections were made using the conservative
case ($q_\mathrm{conf}>1$) of the visual inspections.
The surface density of tracers using \texttt{redmonster}
is increased by $3.5\%$ relative to those used in the projections.  Because the measurements
of cosmological parameters are Poisson-limited, \texttt{redmonster} provides an additional
2\% margin on achieving the cosmological precision expected from the eBOSS LRG sample.



\subsection{Galaxy redshift precision and accuracy} \label{sec:precision}

\begin{figure}
\plotone{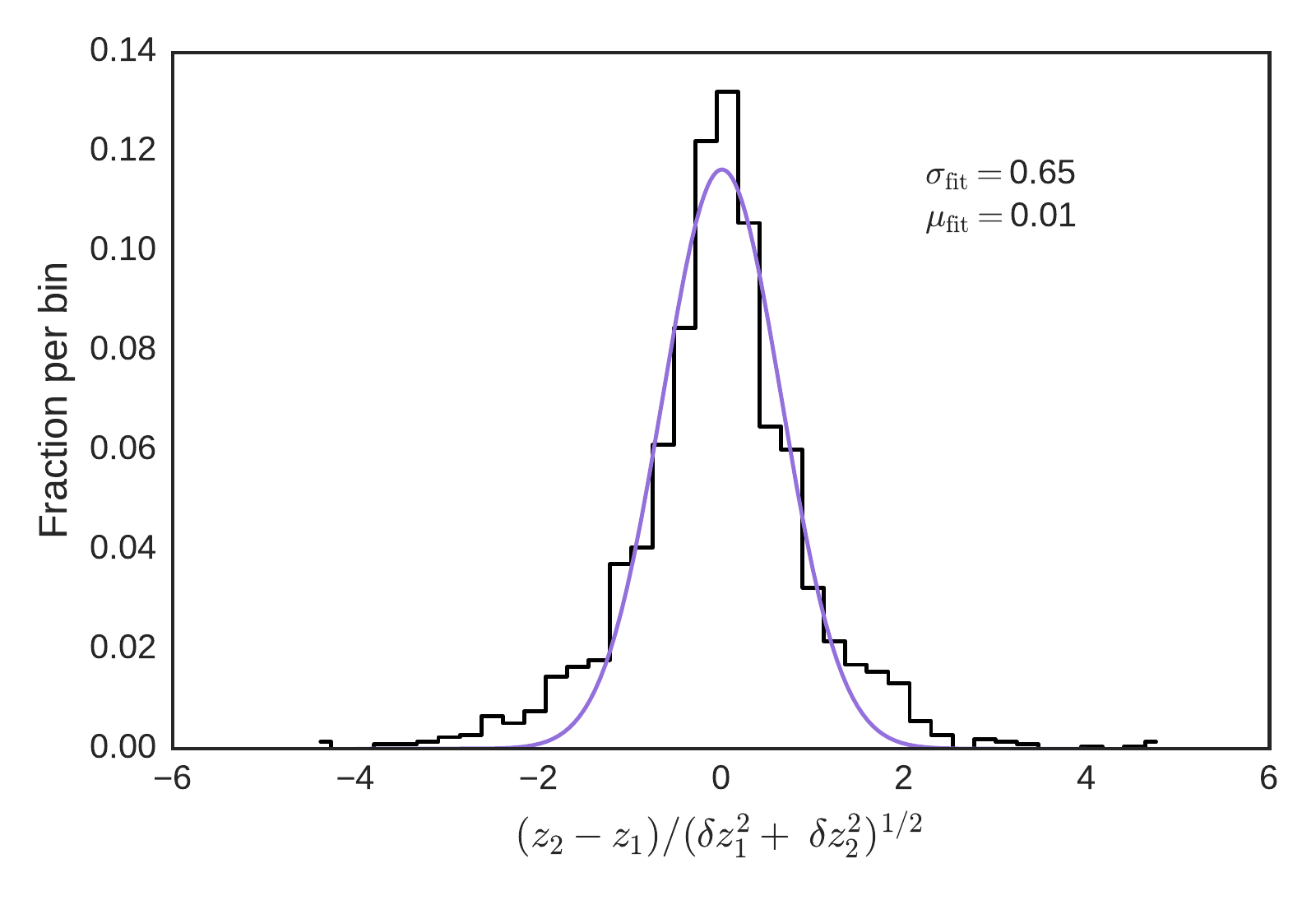}
\caption{Histogram of redshift differences of eBOSS LRG targets on extra-deep plates that
have had observations split into spectra of typical eBOSS depth, scaled by the quadrature
sum of the statistical error estimates of each split.  Over-plotted is the best-fit Gaussian model,
with a standard deviation of $\sigma=0.65$ and a mean of $\mu=0.01$.\\
\label{fig:reobs_errors}}
\end{figure}

Redshift errors are calculated from the curvature of the $\chi^2(z)$ function in the vicinity of the
value that is used to determine the best-fit redshift measurement. To assess the precision of these statistical
error estimates, we used
the same repeat spectra as those used to assess catastrophic failure rates.
We scaled the
redshift difference between the two observations by the quadrature sum of the error estimates from the
spectra from each observation. We then assessed the full distribution of the velocity
differences and fit it with a
Gaussian function. If the estimated errors accounted for all the statistical uncertainty, the fit
would have a dispersion parameter of unity and a mean of zero. Figure~\ref{fig:reobs_errors} shows the
results of this analysis.  The fitted dispersion is $\sigma=0.65$ and the mean is $\mu=0.01$.
Thus, redshift errors are overestimated by $\sim$54\%,
meaning that redshift estimates are more precise than reported.
A similar analysis performed on the \texttt{spectro1d} reductions of the SDSS-III CMASS
(for ``constant mass") galaxy sample ($0.4\lesssim z \lesssim 0.7$) in \citet{bol2012} resulted
in a fitted dispersion parameter of $\sigma=1.19$.

Next, we examine the statistical redshift error~distributions as a function of
median S/N in the SDSS $r$-, $i$-, and $z$-bands, and find a weak anti-correlation.  This is as expected,
and is consistent with previous SDSS data sets.  A summary of these statistics is given in
Table~\ref{tab:redshift_errors_sigma}. Finally, for all LRG targets, we also compute the
distribution of estimated redshift errors as a function of redshift.
A summary of the statistics is given in Table~\ref{tab:redshift_errors_z}.
In all cases, typical errors are a few tens of km~s$^{-1}$.
They should be reduced by an additional factor of 1.54 to reflect
the statistical scatter displayed in Figure~\ref{fig:reobs_errors}. These errors are well below the 300
km~s$^{-1}$ redshift precision requirement of the eBOSS galaxy large-scale structure
and redshift space distortion science analyses.


\begin{deluxetable}{cccc}
\tabletypesize{\footnotesize}
\tablecolumns{4}
\tablewidth{0pt}
\tablecaption{\label{tab:redshift_errors_sigma} Mean and standard deviation of estimated statistical redshift error in
several S/N bins in SDSS $r$-, $i$-, and $z$-bands.  Values are given in km~s$^{-1}$.}
\tablehead{
\colhead{SDSS band} & \colhead{S/N range} & \colhead{$\bar{\sigma}_{\mathrm{v}}$} &
\colhead{Var$(\sigma_{\mathrm{v}})^{1/2}$}
}
\startdata
$r$ & $1.0<$ S/N $<1.5$ & 34.50 & 10.41\\
$r$ & $1.5<$ S/N $<2.0$ & 28.10 & 8.83\\
$r$ & $2.0<$ S/N $<2.5$ & 22.38 & 7.99\\
$r$ & $2.5<$ S/N $<3.0$ & 18.59 & 15.25\\
$i$ & $1.0<$ S/N $<1.5$ & 52.18 & 17.40\\
$i$ & $1.5<$ S/N $<2.0$ & 46.50 & 14.80\\
$i$ & $2.0<$ S/N $<2.5$ & 41.56 & 12.52\\
$i$ & $2.5<$ S/N $<3.0$ & 36.76 & 9.97\\
$i$ & $3.0<$ S/N $<3.5$ & 32.89 & 8.88\\
$i$ & $3.5<$ S/N $<4.0$ & 29.56 & 8.48\\
$z$ & $1.0<$ S/N $<1.5$ & 51.51 & 16.88\\
$z$ & $1.5<$ S/N $<2.0$ & 46.10 & 13.99\\
$z$ & $2.0<$ S/N $<2.5$ & 40.38 & 10.96\\
$z$ & $2.5<$ S/N $<3.0$ & 35.68 & 9.42\\
$z$ & $3.0<$ S/N $<3.5$ & 32.02 & 8.49\\
$z$ & $3.5<$ S/N $<4.0$ & 28.93 & 7.38\\
\enddata
\end{deluxetable}

\begin{deluxetable}{ccc}
\tabletypesize{\footnotesize}
\tablecolumns{3}
\tablewidth{0pt}
\tablecaption{\label{tab:redshift_errors_z} Mean and standard deviation of estimated statistical redshift error in several redshift bins.
Values are given in km s$^{-1}$.}
\tablehead{
\colhead{Redshift range} & \colhead{$\bar{\sigma}_{\mathrm{v}}$} & \colhead{Var$(\sigma_{\mathrm{v}})^{1/2}$}
}
\startdata
$0.6<z<0.7$ & 37.36 & 12.24\\
$0.7<z<0.8$ & 38.69 & 13.21\\
$0.8<z<0.9$ & 41.70 & 15.23\\
$0.9<z<1.0$ & 45.22 & 17.33
\enddata
\end{deluxetable}


\subsection{Composite spectra and distribution
of galaxy parameters} \label{sec:composites}

A primary advantage of \texttt{redmonster} over PCA-based redshift classification techniques
is the simple manner in which the best-fitting template can be translated to physical parameters.
In this section, we briefly discuss two types
of analyses made possible by this parameterization.  We
seek only to demonstrate these possibilities and defer analysis of the results to a later work.

Stacking large numbers of spectra has become a widely-used technique in
extragalactic physics and cosmology.
We derive composite spectra of high signal-to-noise ratios to enable analysis
of the quality of our templates in relation to the true spectral features
in eBOSS galaxies.  Previous work (e.g., \citealp{eis2003}) concentrated
on analyzing and averaging spectra based on observed quantities, such as color or magnitude.
The physicality of the \texttt{redmonster} parameterization allows us to separate and bin our galaxy sample
based on quantities such as age, velocity dispersion, emission line ratio and strength,
metallicity, and star formation history (SFH), as determined by the best-fitting template.

\begin{figure*}
\plotone{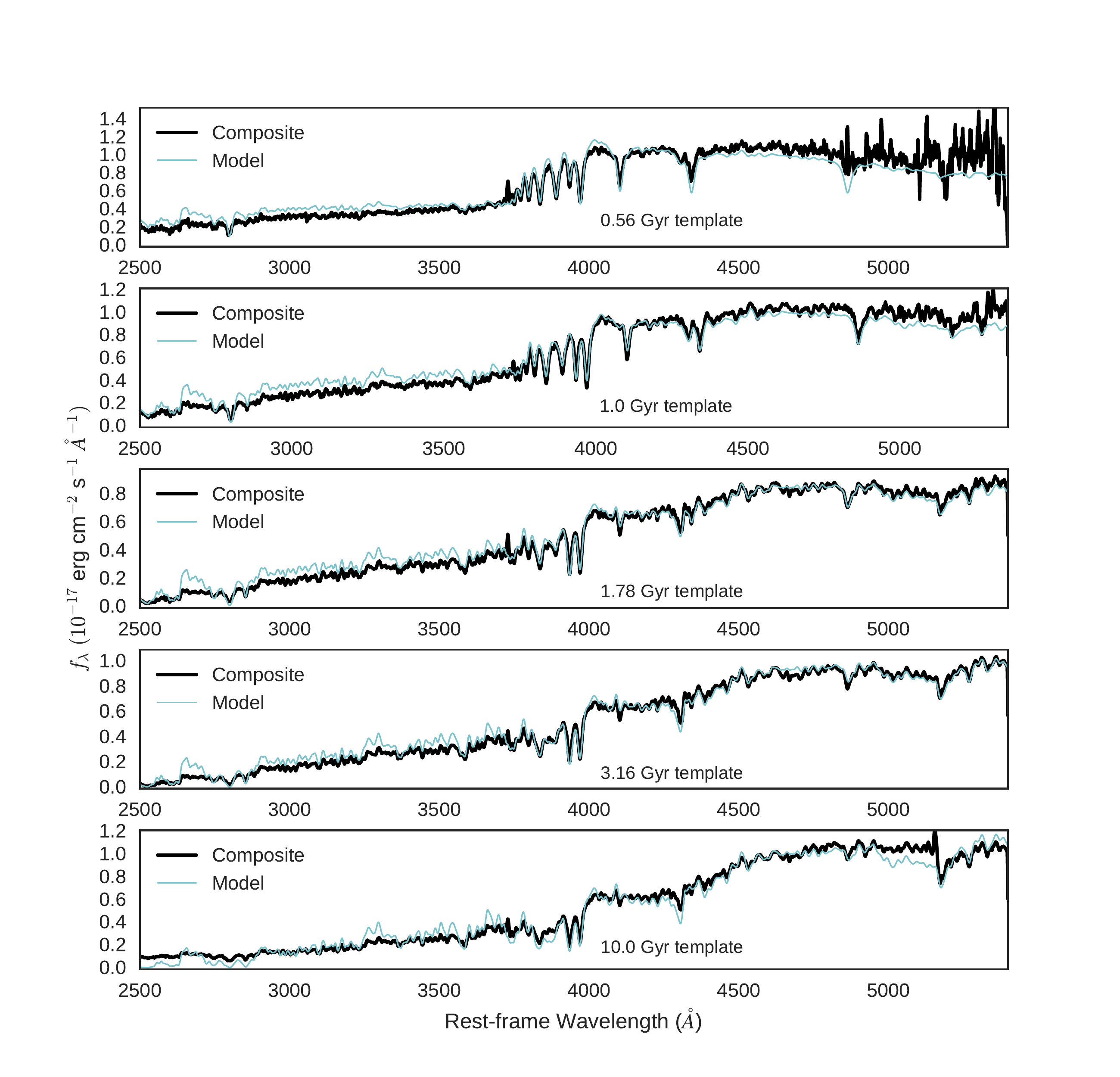}
\caption{Composite spectra for five ages as estimated by \texttt{redmonster}.  The data are
shown in black, and the template of corresponding age is shown in teal.  Here, age is the
only free parameter; thus, one should not over-interpret mismatches between
composites and models.\\
\label{fig:comps}}
\end{figure*}

We first binned a subset of the eBOSS galaxy sample based on the age of the template that produced
the best-fit model. Objects were chosen with best-fitting templates of zero line flux to
identify a passively-evolving sample.
To derive meaningful composites, these spectra then must be normalized.
Due to the relatively low signal-to-noise of eBOSS spectra, we cropped the noisy blue- and red-end
of each spectrum, and scaled the spectrum such that the median of the pixels
in the wavelength range $4000<\lambda<9000$ is unity.
We scaled the template corresponding to each bin to fit the composite spectrum.
Example composite spectra for observed spectra best fit by the 0.56~Gyr, 1.0~Gyr, 1.78~Gyr,
3.16~Gyr, and 5.62~Gyr
templates are shown in Figure~\ref{fig:comps}; they contain 412, 1,486, 14,739,
22,288, and 11,134 unique
galaxies, respectively.  We stress that these composites are selected only by template age;
metallicity is assumed to be solar \citep{asp2009} and velocity dispersion is assumed to
be 250~km~s$^{-1}$.  An example of fitting similar composites over
the wavelength range $0.4-0.8\mu m$
while allowing metallicity and velocity dispersion to vary is given in \citet{con2014}.
These eBOSS data will allow a similar analysis to be extended to shorter wavelengths.

In general, the templates better describe the continuum in the red than the blue.
The templates systematically over-estimate flux density
between $\sim 2500$~\AA\ and $\sim 3600$~\AA.
All three templates have a broad feature at $\sim 2700$~\AA\ that is exaggerated
relative to the composite spectra, likely due to missing atomic opacities in the models.
Discrepancies in the continuum may be due to the effects of dust attenuation.  In the core
fitting algorithm, these effects can be accounted for by the polynomial nuisance vectors;
the models in this figure are scaled to the composite spectra without any polynomial terms,
allowing the effects of dust in the data to appear as shortcomings in the templates.
On the other hand, the models are able to reproduce the observed behavior of the narrow-band features.
All five composite spectra clearly display lines from the Mg~II doublet (2796~\AA\ and 2803~\AA),
Ca~H\&K~(3934~\AA\ and 3968~\AA), H$\delta$~(4103~\AA), G-band~(4307~\AA), and
H$\beta$~(4863~\AA) that are well fit by the templates.  The 0.56~Gyr and 1.0~Gyr composite spectra have a
well-fit H$\gamma$ line~(4341~\AA) and more prominent Balmer features, as expected from a younger
stellar population.
Additionally, an Mg~I line at 2852~\AA, a band of Mg~I absorption
just blueward of Ca~H\&K, and an Mg~I line at 5175~\AA\ become more prominent
at older stellar populations.

\begin{figure}
\plotone{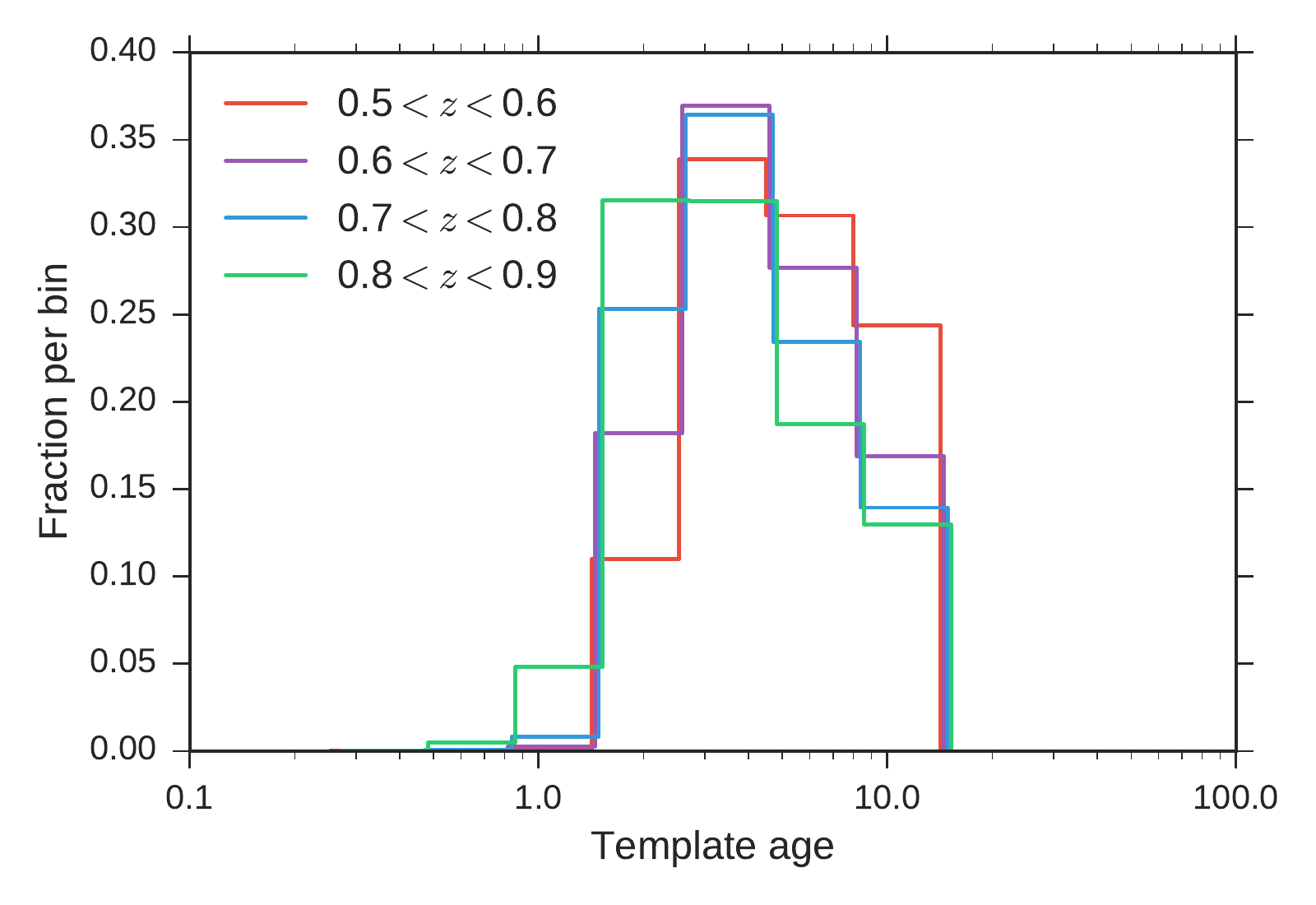}
\caption{Distributions of the age of the template component in the best-fit model to each spectrum
in four redshift slices.\\
\label{fig:ages_histos}}
\end{figure}

Next, we evaluate the distribution of the physical parameters across the galaxy sample.  These distribitions
are informative of both the accuracy of the spectral features in the templates and the targeting
completeness and selection bias of the survey.  As an example, we again consider galaxy
template age.  We binned our sample into four redshift bins
and show the distribution of the ages of the best-fit templates in
Figure~\ref{fig:ages_histos}.  The mean values of
the four redshift bins, $0.5<z<0.6$, $0.6<z<0.7$, $0.7<z<0.8$, and $0.8<z<0.9$, are
4.9 Gyr, 4.3 Gyr, 3.9 Gyr, and 3.7 Gyr, respectively.
Assuming a $\Lambda$CDM cosmology with $k=0$, $H_0=67.8$ km s$^{-1}$ Mpc$^{-1}$,
$\Omega_\mathrm{m}=0.308$, and $w_\Lambda=-1$ \citep{pla2015}, the age of the universe at redshifts
$z=0.56$, $z=0.65$, $z=0.75$, and $z=0.84$, the sample mean in each bin, is 8.18 Gyr, 7.61 Gyr,
7.03 Gyr, and 6.57 Gyr, respectively.  A comparison with the median template age in each
bin reveals a galaxy sample that ages more slowly than the universe.  Due to not allowing
metallicity and abundance patterns to be fit as free parameters (as the templates in this paper
use only solar metallicity), it is likely not possible to meaningfully interpret the ages.  However,
a more careful analysis could be used to investigate targeting selection bias in the survey.

\section{Summary and Conclusion} \label{sec:conclusion}

We have described the \redmonster~software that provides automated redshift measurement and
spectral classification and its performance on the SDSS-IV eBOSS LRG sample, comprising
99,449 spectra.  This software provides a new algorithm and new sets of templates that
restrict all spectral fitting to only physically-motivated models.
The advantages over the current algorithm include robustness against unphysical solutions
likely to arise for low signal-to-noise spectra (particularly in the presence of imperfect sky-subtraction),
determination of joint likelihood functions over redshift and physical parameters, and custom
configurability of spectroscopic templates for different target classes.

The redshift success rate
of the \redmonster~software on eBOSS LRGs is 90.5\%, meeting the eBOSS scientific requirement
of 90\% and providing a significant improvement over the previous
redshift pipeline, \texttt{spectro1d}.  The improvement translates to a 23.9\% increase in the surface
density of tracers that can now be used to constrain cosmology through clustering measurements.
We have shown catastrophic failure rates for \texttt{redmonster} of 0.98\%,
in agreement with the scientific requirements of $<1$\%.
The software also provides robust estimates of statistical redshift
errors that are Gaussian distributed, typically a few tens of km~s$^{-1}$,
well below the specified maximum of 300 km~s$^{-1}$.

Looking forward, using $\Delta\chi_\mathrm{threshold}^2=0.0015$ would give
\texttt{redmonster} a completeness of 95.7\% and a catastrophic failure rate
of 1.9\%, which would very nearly meet the DESI science requirements
of at least 95\% completeness and a maximum of 5\% catastrophic failures on eBOSS data.
The raw S/N in DESI will be comparable to that in eBOSS, though the image quality in eBOSS
two-dimensional spectra is
degraded relative to the image quality we expect in
the bench-mounted DESI system.
eBOSS also shows a failure rate that increases to $\sim25\%$ near the edges of the focal
plane.  This is imperfect optics towards the edges of the spectrographs,
which will be less significant in DESI.
Therefore, we expect improved \texttt{redmonster} performance
on the more well-behaved DESI spectra.

Development work is ongoing for eBOSS, both in the calibration and extraction
of spectra and on \redmonster\ itself.  The next priority for \redmonster\ development is
to build and test templates for ELG and quasar spectra.  Additionally, we will incorporate
simultaneous fitting to the individual exposures at their native resolution to remove covariances
between neighboring pixels introduced by the co-adding process.
Subsequent eBOSS data releases will
be accompanied by catalogues of redshift measurements and spectral classifications
produced by \redmonster.

\acknowledgments

Funding for the Sloan Digital Sky Survey IV has been provided by the Alfred P. Sloan Foundation and the Participating Institutions. SDSS-IV acknowledges support and resources from the Center for High-Performance Computing at the University of Utah. The SDSS web site is www.sdss.org.

SDSS-IV is managed by the Astrophysical Research Consortium for the Participating
Institutions of the SDSS Collaboration including the Brazilian Participation Group, Carnegie
Institution for Science, Carnegie Mellon University, the Chilean Participation Group,
Harvard-Smithsonian Center for Astrophysics, Instituto de Astrof\'isica de Canarias,
The Johns Hopkins University, Kavli Institute for the Physics and Mathematics of the
Universe (IPMU) / University of Tokyo, Lawrence Berkeley National Laboratory, Leibniz
Institut f\"ur Astrophysik Potsdam (AIP), Max-Planck-Institut f\"ur Astrophysik (MPA Garching),
Max-Planck-Institut f\"ur Extraterrestrische Physik (MPE), Max-Planck-Institut f\"ur Astronomie
(MPIA Heidelberg), National Astronomical Observatory of China, New Mexico State University,
New York University, University of Notre Dame, Observat\'orio Nacional do Brasil,
The Ohio State University, Pennsylvania State University, Shanghai Astronomical Observatory,
United Kingdom Participation Group, Universidad Nacional Aut\'onoma de M\'exico,
University of Arizona, University of Colorado Boulder, University of Portsmouth,
University of Utah, University of Washington, University of Wisconsin, Vanderbilt University,
and Yale University.

The work of TH, AB, KD, JB, and MV was supported in part by the U.S. Department of
Energy, Office of Science,
Office of High Energy Physics, under Award Numbers DE-SC0010331 and DE-SC0009959,
and that of JN and AP under Award Number DE-SC0007914.

\pagebreak

\appendix

\section{Output files} \label{sec:output}

The \texttt{redmonster} software generates two output files for each input set of spectra
and summary files of all objects processed by the software.  The ``\texttt{redmonsterAll}'' file is the
primary output and contains all redshifts, spectral classifications, parameter estimates,
and models, and is described in \S\ref{sec:redmonsterfile}.  The ``\texttt{chi2arr}"
file is an optional output containing the entire $\chi^2(\vec{P},z)$ surface for a
template class and is described in \S\ref{sec:chi2arrfile}.  A more detailed description of these files
can be found in the online documentation\footnote{https://data.sdss.org/datamodel/files/REDMONSTER\_SPECTRO\_REDUX}.

\subsubsection{\texttt{redmonster} files} \label{sec:redmonsterfile}

The ``\texttt{redmonsterAll}" file is the primary output of the software. This file contains all
redshift and parameter measurements, spectral classifications, and the best fit model
for each object.  This output is an uncompressed FITS file with all
relevant information in the primary HDU header and first BIN table. The
contents of the BIN table are detailed in Table~\ref{tab:redmonsterfile}. It has the naming scheme
\textit{redmonsterAll-vvvvvv.fits}, where \textit{vvvvvv} is the version of the reduction used.
This file is the parallel to the \texttt{spAll} file produced by \texttt{spectro1d} in BOSS and eBOSS.

Additionally, a batch file is created for each batch of spectra processed. It is the parallel to \texttt{spectro1d}'s
\texttt{spZall} for an SDSS plate in BOSS and eBOSS, and is named
\texttt{redmonster-pppp-mmmmm.fits}, where \texttt{pppp} is the plate number and \texttt{mmmmm}
is the MJD. The file contains a primary header,
a binary extension (similar in format to that of \texttt{redmonsterAll}) containing the best
five redshifts and classifications for each fiber, and an imageHDU containing
a two-dimensional image of the five best-fit models for all fibers. The size of this file is approximately 20 MB
for 1000 spectra of $\sim 4600$ pixels each.

\subsubsection{\texttt{chi2arr} file} \label{sec:chi2arrfile}

The software also has the ability to write the full $\chi^2(\vec{P},z)$ surfaces for each template
class to an output file.  These are also uncompressed FITS files. The primary HDU
contains a multi-dimensional image of all $\chi^2$ values for a single template class for each spectrum.
These files follow the naming scheme \textit{chi2arr-ttttttt-vvvvvv.fits}, where \textit{tttttt}
is the name of the template class and \textit{vvvvvv} is the version
of the reduction used.  The primary HDU contains a multi-dimensional array of the full
$\chi^2$ surface for that template class.  These files are written per batch of spectra processed.
A summary file similar to \texttt{redmonsterAll} does not exist due to the large size
of these files (often several GB per 1000 spectra).

\begin{deluxetable}{ll}
\tabletypesize{\footnotesize}
\tablecolumns{2}
\tablewidth{0pt}
\tablecaption{\label{tab:redmonsterfile} \texttt{redmonsterAll} file binary table contents}
\tablehead{
\colhead{Name} & \colhead{Description}
}
\startdata
\texttt{FIBERID} & \textit{spPlate} fiber number (0-based) for each object\\
\texttt{PLATE} & \textit{spPlate} plate number for each object\\
\texttt{MJD} & \textit{spPlate} MJD for each object\\
\texttt{DOF} & Degrees of freedom used to calculate $\chi_\mathrm{red}^2$\\
\texttt{BOSS\_TARGET1} & BOSS targeting bit\\
\texttt{EBOSS\_TARGET0} & SEQUELS targeting bit\\
\texttt{EBOSS\_TARGET1} & EBOSS targeting bit\\
\texttt{Z} & Best redshift (least $\chi_\mathrm{red}^2$)\\
\texttt{Z\_ERR} & 1-$\sigma$ error associated with Z\\
\texttt{CLASS} & Object type classification\\
\texttt{SUBCLASS} & Best-fit template parameters\\
\texttt{FNAME} & Full name of \textit{ndArch} file of template\\
\texttt{MINVECTOR} & Coordinates of best-fit template in \textit{ndArch} file\\
\texttt{MINRCHI2} & $\chi_\mathrm{red}^2$ value of fit\\
\texttt{NPOLY} & Number of additive polynomials used\\
\texttt{NPIXSTEP} & Pixel step size used\\
\texttt{THETA} & Coefficients of template and polynomial terms in fit\\
\texttt{ZWARNING} & \texttt{ZWARNING} flags\\
\texttt{RCHI2DIFF} & $\Delta\chi_\mathrm{red}^2$ between first- and second-best fits\\
\texttt{CHI2NULL} & $\chi_\mathrm{null}^2$ value for each spectrum\\
\texttt{SN2DATA} & (S/N)$^2$ of each spectrum
\enddata
\tablecomments{Each SDSS plate's reduction file contains
the best five redshifts, errors, and classifications (\texttt{Z1}, \texttt{Z\_ERR1}, \texttt{CLASS1}, etc.).
}
\end{deluxetable}

\clearpage


\clearpage


\end{document}